\documentstyle[aps,eqsecnum,epsfig]{revtex}

\newcommand{\dddot}[1]{\stackrel{...}{#1}}
\newcommand{\dss}{\displaystyle}
\newcommand{\pkt}{\; .}
\newcommand{\kma}{\; ,}
\newcommand{\tr}{{\;\rm tr \; }}

\newcommand{\re}{{\rm Re }\,}
\newcommand{\im}{{\rm Im }\,}

\newcommand{\dslash}{ \slash\hspace{-2.3mm} \partial}

\newcommand{\intp}{\int\frac{d^3p}{(2\pi)^3 2\Ep\,}}
\newcommand{\intpp}{\int\frac{d^3p'}{(2\pi)^3 2\Epp\,}}
\newcommand{\intppp}{\int\frac{d^3p''}{(2\pi)^3 2\Eppp\,}}
\newcommand{\bfp}{{\bf p}}

\newcommand{\bfq}{{\bf q}}
\newcommand{\bfx}{{\bf x}}
\newcommand{\phit}{{\tilde\phi}}
\newcommand{\sigmat}{{\tilde\sigma}}
\newcommand{\Sigmat}{{\tilde\Sigma}}
\newcommand{\bgamma}{\mbox{\boldmath{$\gamma$}\unboldmath}}
\newcommand{\disc}{{\rm disc}}
\newcommand{\bsigma}{\mbox{\boldmath{$\sigma$}\unboldmath}}
\newcommand{\bSigma}{\mbox{\boldmath{$\Sigma$}\unboldmath}}
\newcommand{\be}{\begin{equation}}
\newcommand{\ee}{\end{equation}}
\newcommand{\bea}{\begin{eqnarray}}
\newcommand{\eea}{\end{eqnarray}}

\newcommand{\Ep}{E_{p}}
\newcommand{\Epp}{E_{p'}}
\newcommand{\Eppp}{E_{p''}}
\tighten
\begin{document}
\preprint{DO-TH-99/11; LPTHE-99/19} 
\title{\bf  INITIAL TIME SINGULARITIES IN NON-EQUILIBRIUM EVOLUTION OF
CONDENSATES AND THEIR RESOLUTION IN THE LINEARIZED APPROXIMATION }  
\author{{\bf J. Baacke$^{(a,c)}$, D. Boyanovsky$^{(b,c)}$, H. J. de Vega$^{(c,b)}$
}} \address { (a) Institut f\"ur Physik, Universit\"at Dortmund,
D-44221 Dortmund, GERMANY \\
(b) Department of Physics and Astronomy, 
University of Pittsburgh, Pittsburgh, PA 15260 USA\\ (c)
LPTHE, Universit\'e Pierre
et Marie Curie (Paris VI) et Denis Diderot (Paris VII), Tour 16, 1er. \'etage,
4, Place Jussieu 75252 Paris, Cedex 05, FRANCE.\footnote{Laboratoire
Associ\'{e} au CNRS UMR 7589.}} \date{\today} \maketitle 
\begin{abstract}
The real time non-equilibrium evolution of condensates in field theory
requires an initial value problem specifying an initial 
quantum state or density matrix. Arbitrary specifications of the
initial quantum state (pure or mixed) results 
in initial time singularities which are not removed by the usual
renormalization counterterms. We study the initial time
singularities in the linearized equation of motion for the scalar condensate 
in a renormalizable  Yukawa theory in $ 3+1 $ dimensions. In this 
renormalizable theory the initial time singularities are enhanced. 
We present a consistent method for removing these initial time
singularities by specifying initial states where the distribution of high
energy quanta is determined by the initial conditions and  the interaction
effects. This is done through a  Bogoliubov transformation which is
consistently obtained in a perturbative expansion.  The usual
renormalization counterterms and
the proper choice of the Bogoliubov coefficients lead to a singularity
free evolution equation. We establish the relationship between the
evolution equations in the linearized approximation and linear
response theory. It is found that only a very specific form of the
external source for linear response leads to a real time evolution
equation which is singularity free. We focus on  the evolution
of {\bf spatially inhomogeneous} scalar condensates by implementing 
the initial state preparation via a Bogoliubov transformation 
up to  one-loop. As a concrete application, the evolution equation for
an inhomogenous condensate is solved analytically and the results are
carefully analyzed. Symmetry breaking by initial quantum states is discussed.  
\end{abstract} 

\section{Introduction}
The study of the real time dynamics and the evolution of non-equilibrium
quantum states has now become ubiquitous in cosmology and
high/intermediate energy physics. In cosmology the real time evolution
of expectation values of quantum fields is a necessary component of a
microscopic description of the inflationary dynamics and the subsequent
hot FRW stage (big bang) seeking to give a realistic description of
the early universe and the physical processes originated there. 
In the physics of heavy ion collisions a very active program seeks to
establish potential experimental signatures from possible non-equilibrium 
stages of the evolution of the quark-gluon and chiral phase
transitions\cite{QCD}. In cosmology a program that incorporates
consistently the non-equilibrium evolution of initial quantum states
or density matrices of thermal or non-thermal origin including
renormalization and backreaction effects had been pursued vigorously
during the last few years\cite{boycosmo,coom,baacke:1997,ramsey}. In
high/intermediate energy the possibility of studying the quark-gluon
plasma and chiral phase transition at the forthcoming
ultrarelativistic heavy ion colliders (RHIC and LHC) has 
motivated a substantial effort to study out of equilibrium dynamics during 
phase transitions. In particular the formation of coherent pion
domains\cite{dcc}, the evolution of non-equilibrium initial 
 density matrices and states of high energy density\cite{boydcc}, and
isospin condensates\cite{vautherin}. Non-perturbative techniques had
been developed to study consistently non-equilibrium dynamics of
quantum field theories\cite{largeN,boycosmo} and current computational
facilities allow the possibility of studying the non-equilibrium
dynamics of non-linear, inhomogeneous configurations in quantum field
theories\cite{boycoopveg} including gauge
theories\cite{kluger,baackegauge}, for which recent lattice
simulations of non-equilibrium gauge field theories with topological
excitations had recently 
been reported\cite{aarts}. 

The real time evolution of either density matrices or pure states, or
alternatively of matrix elements must be set up as an {\em initial
value} problem, either by specifying the initial state or by providing
the Cauchy data (expectation values and their derivatives) typically
on space-like hypersurfaces. Once this initial value problem has been
set up at some initial time, the real-time evolution of the
expectation values or other matrix elements  can be studied either
analytically in the case of small amplitudes\cite{boyinho,boyareheat}
or numerically in the case of large amplitude
configurations\cite{largeN,boydcc,boycoopveg,baackegauge,aarts}
(Although analytic expressions are available in some extent).  

An important but largely unnoticed subtlety arises in these situations
in that besides the usual ultraviolet divergences associated with masses,
couplings and wave-function renormalizations there appear {\bf initial time
singularities}\cite{Baacke:1998a,Baacke:1998c}. The physical reason for
these initial time singularities can be understood as follows: the initial
state (either  pure or mixed) is typically chosen to reflect some
physical description but generally is either  some initial pure
excited state with free field quanta or a thermal density matrix for
free field theory. The choice of the initial state (including the
field expectation value and its time derivative) has been essentially
arbitrary and in particular  independent of the field hamiltonian.
The time evolution with the interacting Hamiltonian suddenly couples 
at the initial time the infinite number of degrees of freedom of the
theory, redistributing the spectral densities. In the case 
in which the underlying theory is renormalizable this redistribution
of the spectral densities results in a divergent response.  Such
effect is also present on systems with a finite number of degrees of
freedom but it is then finite. 

The consideration of singularities associated with setting up initial
conditions in a quantum field theory has been addressed originally by 
Stueckelberg\cite{stu}, the similarity with sharp boundary conditions
in a Euclidean formulation has been studied by Symanzik\cite{sym} and
has since found different possible
solutions\cite{stu,coom,Baacke:1998a,Baacke:1998c,ruso}. A very appealing 
method to prepare initial states  that lead to evolution equations
{\em without} initial time singularities has been recently
advocated\cite{Baacke:1998a,Baacke:1998c} for self-consistent
real-time evolution. This method consists in defining an initial state as a 
Bogoliubov transformation of the initial states in a free field
theory. The Bogoliubov transformation is chosen  to cancel the initial time
singularities. The advantage of this method is that it is physically
transparent and can be implemented both for small amplitude, i.e. the
linearized problem, as well as for the large amplitude case which must
be necessarily studied numerically. 

In this article we focus on studying these initial time singularities
and their resolution via the method of a Bogoliubov transformed
initial state in a {\em renormalizable} theory in the case of small
amplitudes of the scalar condensate. This case allows to obtain the
evolution equations in a linearized approximation with an analytical
solution to the evolution. Furthermore, we establish the correspondence 
between this method and linear response theory for the case of linearized
equations of motion of condensates. An important corollary of this
correspondence is that only very specific choices of the external
source in the linear response approach lead to a singularity free
initial value problem. We choose to study the initial value
problem for the evolution of a scalar field condensate in a Yukawa
theory in $3+1$ dimensions both for homogeneous as well as for
inhomogeneous condensates. Whereas in\cite{Baacke:1998a,Baacke:1998c}
the homogeneous case has been studied in a self-consistent manner and
the linearized approximation has been extracted from it, the {\em
inhomogeneous} case has not been studied, and hence we devote our
attention mainly to this important case. In a 
renormalizable theory the initial time singularities are enhanced and new
infinities  associated with the preparation of the initial state emerge, this
situation is highlighted  in the renormalizable  Yukawa theory which
is the focus of our study. 

{\bf Main Results:} i) The main results of our study can be summarized
as follows: a definite proposal to set up the initial value problem in
renormalizable quantum field theories based on an initially Bogoliubov
transformed state. In order to eliminate the initial time
singularities, the Bogoliubov coefficients are constrained to vanish in a
particular manner for high momentum modes. In particular the
Bogoliubov coefficients for a mode of momentum $p$ must be chosen such
that the $1/p~,1/p^3$ contributions to these coefficients  are {\bf uniquely
fixed} by the initial data, the coupling and the mass
[sec. IV]. Choices of Bogoliubov coefficients that differ by
contributions of higher order in $1/p$ define different initial
states, all of them {\bf free of initial time singularities}. Thus, the
time evolution of an initial state in quantum field theory  is free of
initial time singularities provided that high energy distribution of
quanta of the initial state is specified in a very precise  manner.  

This method is implemented consistently in the
perturbative expansion and in combination with the usual
renormalization of mass, wave-function and coupling leads to a
real-time evolution free of ultraviolet and initial time
singularities.  ii) As an example we study  
the real-time evolution of an {\bf inhomogeneous} scalar condensate in
the Yukawa theory, both in the case in which the scalar is heavy and
can decay into fermion-antifermion pairs, and in the case in which the
scalar is light and cannot decay into fermion pairs. Here we provide a
detailed analysis of the real time evolution of an inhomogeneous
scalar condensate corresponding to a spherical wave. 

The article is organized as follows: in section II we obtain the equations
of motion for a scalar condensate. In section III we analyze the
ultraviolet and initial time singularities. For simplicity we present first
the case of a homogeneous scalar condensate. Section IV introduces the
Bogoliubov transformed initial state in the case of a homogeneous
condensate, 
 discusses in detail the choice of the Bogoliubov coefficients that lead to
an evolution free of initial time singularities and presents the
singularity free real-time equations of motion for the homogeneous
case. In section V we establish a relation between the initial value
problem in the linearized approximation and linear response and
discuss the constraints on the external sources 
that lead to a well defined initial value problem free of singularities.
 In section VI we extend the treatment to
the case of inhomogeneous scalar condensates, obtain the corresponding 
inhomogeneous Bogoliubov transformation consistently in perturbation
theory and the equations of motion free of singularities to one-loop
order. In section VII we obtain an analytic solution of the real-time
equations of 
evolution for an inhomogeneous scalar condensate. We also provide a
numerical analysis of the solution and discuss its main features.  
The conclusions summarize our work and discusses the potential
applications of the methods presented. 

The Bogoliubov transformation of the tadpole and self-energy diagrams
is presented in the Appendices A-C. Furthermore, Appendix D
establishes several sum rules on the spectral densities.  

\section{Linearized equations of motion for condensates:}
\setcounter{equation}{0}
Although the initial time singularities that will be discussed in this
article are generic features of initial 
value problems in field theory, they are highlighted in renormalizable
theories. Therefore we choose to discuss these singularities 
and their resolution in a Yukawa theory in $3+1$ dimensions. The focus
of this article is to understand the problems 
of setting up an initial value problem to describe the non-equilibrium
evolution of condensates or  field expectation values in the 
linearized (small amplitude) approximation. Furthermore, we compare with an
alternative formulation based on linear response.  

We consider a massive scalar field $ \Phi(x) $ coupled to a massive
Dirac field $ \psi(x) $ in a Yukawa model specified by the Lagrangian density
\be
{\cal L}(\Phi,\psi,\bar\psi)=
\frac{1}{2}\partial_\mu\Phi(x)  \; \partial^\mu\Phi(x)
+\frac{1}{2}M^2 \; \Phi^2(x)+ \bar\psi(x)\left[i \dslash + m
+g  \; \Phi(x)\right] \psi(x) \label{lagra}
\pkt\ee
We study the time 
evolution of the expectation value of the scalar field 
via the real time generating functional in terms of a path integral
defined on a contour in complex time (CTP)\cite{ctp,keldysh}. The
effective Lagrangian that enters in the contour path integral is 
\bea
{\cal L}_{eff}= {\cal L}(\Phi^+,\psi^+,\bar\psi^+)-{\cal
L}(\Phi^-,\psi^-,\bar\psi^-) \nonumber
\kma
\eea
where the $\pm$ labels on the fields refer to the forward ($+$) and
backward ($-$) branches corresponding to the 
forward and backward time evolution of the initially prepared density
matrix. We now follow the procedure of  
references\cite{boyinho,boyareheat} and use the tadpole method to
obtain the equation of motion for the expectation 
value of the scalar field 
$$
\phi(x) \equiv \; <\Phi(x)> 
$$  
and  write 
\bea
\Phi^{\pm}(x)= \chi^{\pm}(x) + \phi(x) \; \; ; \; \; \langle
\chi^{\pm}(x) \rangle =0 \pkt\nonumber
\eea
We specify the initial data at the  time $ t_0 = 0 $ by giving the
initial condition, 
\bea
\phi({\bf x},0)\equiv \phi({\bf x}) \quad \mbox{and}\quad \dot
\phi({\bf x}, 0)\equiv \dot \phi({\bf x})  
\pkt \nonumber
\eea
Let us consider that the initial density matrix at time $ t_0=0 $ is given by
\bea
\rho(0) = |0\rangle\langle0| \nonumber
\eea
\noindent with $ |0> $ the free field Fock vacuum for the scalar and
fermion fields. For $ t>0 $, 
$ \rho(t)= e^{-iHt} \;  \rho(0) \; e^{iHt} $ where $ H $ is the {\em full}
Hamiltonian. This case is tantamount to considering an  
initial free field vacuum state and switching-on the interaction
suddenly at $t=0$.  

The equation of motion for $\phi({\bf x},t)$ is obtained in a
systematic perturbative expansion by imposing that 
$\langle \chi^{\pm}({\bf x},t) \rangle=0$ to all orders in
perturbation theory. We will restrict our study to the 
case of small amplitudes of the condensate and will obtain the
equations of motion {\em linearized} in $\phi({\bf x},t)$.  
In this linear approximation the self-energy kernel is obtained to any
desired order in a  perturbative expansion in the 
Yukawa coupling but in the state with {\em vanishing condensate}. Thus
assuming that the state with vanishing condensate 
is spatially translational invariant it is convenient to perform a
spatial Fourier transform for the condensate and 
the self-energy kernel\cite{boyinho,boyareheat}. Anticipating
renormalization effects we introduce the renormalized 
field and mass in the Lagrangian before shifting the field by the condensate

\bea
\Phi(x) = \sqrt{Z}_{\phi} \; \Phi_R(x) ~~;~~
Z_{\phi}\;M^2=M^2_R+\delta M^2 \nonumber
\pkt \eea

Since the fermionic fields will be integrated out to obtain the
equation of motion for the expectation value of the scalar field, we
do not introduce the renormalizations associated with the fermionic
fields. We now drop the subscript $ R $ from the renormalized
quantities to avoid cluttering of notation, with the understanding
that the scalar field and its mass are the renormalized ones.  

The equation of motion for the spatial Fourier transform of the condensate 
$$
\phi_{\bf q}(t) \equiv \int d^3x \; e^{-i {\bf q}\cdot{\bf x}}\;
\phi({\bf x}, t)  \; ,
$$
to one loop order is given by (see ref.\cite{boyareheat} for details)
\be
(1+\delta Z )\left[\;\ddot \phi_{\bf q}(t)+q^2\; \phi_{\bf
q}(t)\right] + (M^2 +\delta M^2)\; 
\phi_{\bf q} (t) + \int_0^t dt'\; \Sigma_{\bf q} (t-t')\; \phi_{\bf q}(t') 
- J=0 \label{evolequ}
\kma\ee
with $\delta Z= Z_{\phi}-1$ and
\begin{eqnarray}
&& J  = - i g\; \delta_{\bf q,0} \int \frac{d^3k}{(2\pi)^3} \tr
S^>_{\bf k}(t,t)\kma \label{tadeq} \\ 
&&\Sigma_{\bf q}(t-t')=-ig^2 \int \frac{d^3p}{(2\pi)^3} \tr
\left[ S^>_{\bf p}(t-t') S^<_{{\bf p}-{\bf q}}(t'-t)
-  S^<_{\bf p}(t-t') S^>_{{\bf p}-{\bf q}}(t'-t) \right]
\label{selfenerequi}
\pkt\end{eqnarray}
The contribution $J$ is due to the fermion tadpole, it is usually
absorbed in a constant  shift of the field $ \phi(x) $.
The fermionic Green's functions in the vacuum state are given
by\cite{boyareheat} 
\begin{eqnarray}
&& S_{\bf p}^>(t,t')=-i\int d^3x \; e^{-i{\bf p}\cdot
{\bf x}} \; \langle \psi({\bf x},t) \bar{\psi}({\bf 0},t')
\rangle \nonumber \\
&&\quad\quad\quad =-\frac{i}{2E_p}\left[e^{-iE_p(t-t')}
(\not\!{p}+m)+e^{iE_p(t-t')}\gamma_0
(\not{p}-m)\gamma_0  \right]\;,
\nonumber \\
&& S_{\bf p}^<(t,t')=i\int d^3x \; e^{-i{\bf p}\cdot
{\bf x}} \;  \langle \bar{\psi}({\bf 0},t')  \psi({\bf x},t)
\rangle \nonumber \\
&&\quad\quad\quad = \frac{i}{2E_p}\left[e^{-iE_p(t-t')} (\not\!{p}+m)
+e^{-iE_p(t-t')}\gamma_0 (\not\!{p}-m)\gamma_0  \right]\;,
\nonumber \\ &&\quad E_p=\sqrt{{\bf p}^2+m^2}\nonumber
\pkt\end{eqnarray}

\section{Ultraviolet renormalization and initial time singularities}
\setcounter{equation}{0}

The evolution equation of the type (\ref{evolequ}) contains two types 
of divergences: i) ultraviolet divergences which are removed by the
mass and wave function  renormalizations, ii) initial time singularities.

To illustrate these singularities in a more clear manner we now focus
on the case of  homogeneous condensate, i.e. $ {\bf q}=0 $. We find

\begin{eqnarray}
&&\Sigma_{\bf 0}(t-t')=-2 g^2\intp \frac{8p^2}{2\Ep}\;
\sin[2\Ep(t-t')] \label{selfhomoeq}\kma \\
&&J= - 4\;m\;g\intp \nonumber \pkt
\end{eqnarray}
Whereas $J$ acts as a constant source term and can be absorbed in a
shift of the expectation value $ \phi(x) $, the self-energy 
kernel leads to ultraviolet divergences as can be seen upon
integrating by parts the non-local term in (\ref{evolequ}) three times 

\bea
&&\int_0^t \; dt' \; \Sigma_{\bf 0}(t-t') \;\phi_{\bf 0}(t')=
- g^2\intp\frac{8p^2}{\Ep} \int_0^t dt' \; \sin[2\Ep (t-t')]\;
{\phi}_{\bf 0}(t') =\nonumber  \\ \nonumber
&&- g^2\intp\frac{8p^2}{\Ep} \left\{\frac{1}{2\Ep}\; \phi_{\bf 0}(t)
-\frac{1}{2\Ep}\;\phi_{\bf 0}(0)\cos(2\Ep t) 
-\frac{1}{(2\Ep)^2}\;\dot \phi_{\bf 0}(0) \;\sin2\Ep t \right .
\\ \nonumber && \left. -\frac{1}{(2\Ep)^3}\;\ddot \phi_{\bf 0}(t)
+\frac{1}{(2\Ep)^3} \; \ddot \phi_{\bf 0}(0) \;
\cos2\Ep t+\frac{1}{(2\Ep)^3}\int_0^t dt' \; \cos[2\Ep(t-t')]
\stackrel{\dots}{\phi}_{\bf 0}(t')\right\} 
\pkt\nonumber \eea 
Using dimensional regularization the coefficient of $\phi_{\bf 0}(t)$ becomes
\bea
-\delta M^2 = - g^2\intp\frac{4p^2}{\Ep^2}=
\frac{3 g^2 m^2}{4 \pi ^2}\left[\frac{2}{\epsilon} -\gamma+\frac{1}{3}
+\ln\frac{4\pi \mu^2}{m^2}\right]
\pkt\nonumber \eea
This agrees with the expression obtained by
evaluation the corresponding Feynman graph in $ 4-\epsilon $
dimensions. The renormalization is 
performed here at $ q^2 = 0 $. 

The coefficient proportional to $ \ddot\phi_{\bf 0}(t) $ is the
wave function renormalization which again in dimensional
regularization is given by 
\bea
-\delta Z =  g^2\intp\frac{p^2}{\Ep^4}
=\frac{ g^2}{8 \pi ^2}\left [\frac{2}{\epsilon} -\gamma-\frac{2}{3}
+\ln\frac{4\pi \mu^2}{m^2}\right]
\pkt\nonumber \eea
These ultraviolet renormalizations are cancelled by the mass and wave
function counterterms but there still remain 
singularities arising from the terms that are evaluated at the initial
time $t=0$, these are given by 
\bea
&&\left.\left[\int_0^t\; dt'\;
\Sigma_{\bf 0}(t-t')\;\phi_{\bf 0}(t')\right]\right|_{\rm sing} =\nonumber 
\\ \nonumber
&&- g^2\intp\frac{8p^2}{\Ep} \left\{
-\frac{1}{2\Ep}\;\phi_{\bf 0}(0)\;\cos2\Ep t
-\frac{1}{(2\Ep)^2}\;\dot \phi_{\bf 0}(0)\;\sin2\Ep t \right .
\\ \nonumber && \left. 
+\frac{1}{(2\Ep)^3}\; \ddot \phi_{\bf 0}(0)\;
\cos2\Ep t \; \;\right\} 
\pkt\nonumber \eea 

Obviously this expression is singular as $t\to 0$. 
Simple power counting shows that the coefficients of  $\phi_{\bf 0}(0)$,
$\dot\phi_{\bf 0}(0)$, and $\ddot \phi_{\bf 0}(0)$ diverge as 
$1/t^2, 1/t$, and $\log t $, respectively. At finite $t$ they are
finite due to the oscillatory behavior of the integrand.

The physical reason for these singularities is the following: having
prepared the initial state to be the free 
field Fock vacuum and switching-on the interaction suddenly at 
the initial time $ t=0 $, the interaction 
redistributes the spectral density of fields. The 
scalar field states  overlap with the fermionic continuum
of states and the particles become dressed by the interaction. 
This dressing effect which is responsible for mass, wave
function and coupling renormalizations  occurs suddenly when the
interaction is switched on and is reflected as an initial time
singularity. Obviously these  
short time singularities will be present at finite temperature or
density, and are a consequence of the fact that the the underlying
field theory posses an infinite number of degrees of freedom. 

Our main point in this article is that these initial time
singularities can be removed and an initial value problem can 
be defined consistently and free of singularities by considering an
appropriately chosen initial state that is dressed by the interactions. 
We thus propose to initialize the real time evolution by providing an
initial state that includes the dressing as 
a Bogoliubov transformation from the free field Fock
states. Furthermore we will argue that this construction leads to a 
consistent initial value problem for real-time dynamics and can be
implemented systematically order by order in perturbation theory.  

We now discuss in detail this procedure in the example under
consideration to lowest order in the Yukawa coupling.     

\section{Equations of motion with Bogoliubov transformed states:
homogeneous case}  
\setcounter{equation}{0}
In this section we introduce the Bogoliubov transformed states and show
explicitly how the introduction of these dressed states provides a
solution to the problem of initial time singularities. For the sake of
clarity we study first the homogeneous case ${\bf q}=0$ and postpone to
a later section the generalization to inhomogeneous condensates. 

From the free field Fock vacuum state $|0\rangle$ a Bogoliubov
transformed state is obtained after a unitary transformation 

\bea 
|0_{\flat} \rangle = e^{-Q }|0\rangle
\pkt\nonumber
\eea
with $Q$ a unitary operator. Bogoliubov transformed operators  are defined
via 
\bea
 {\cal O}_{\flat}=\exp(-Q)\; {\cal O} \; \exp(Q) 
\kma\nonumber \eea
\noindent therefore if the vacuum state $|0>$ is annihilated by the
destruction operators, the Bogoliubov transformed annihilation operators
annihilate the state $|0_{\flat} \rangle$. 

To lowest order in the Yukawa coupling the Bogoliubov transformation  that 
required to cancel the initial time singularities  only involve fermionic
fields. Only when scalar contributions in higher order corrections arise
there will be a need to introduce the Bogoliubov transformation for
scalar fields.  

Therefore we introduce the antihermitian operator $Q$ that generates the
unitary Bogoliubov transformation 

\bea
Q=\sum_s\intp \beta_{ps}
\left[d^\dagger(-\bfp,s)b^\dagger(\bfp,s)\; e^{i\delta_{ps}}
-b(\bfp,s)d(-\bfp,s)\; e^{-i\delta_{ps}}\right]
\pkt\nonumber\eea

This generator of Bogoliubov transformations illuminates at once the nature
of the Bogoliubov transformed initial states. Acting on the free field Fock
vacuum state, the Bogoliubov transformation leads to a state that is a 
linear combination of particle-antiparticle pairs ot total zero momentum.
The fact that the total momentum of these pairs is zero is of course 
a result of the fact that the scalar condensate is homogeneous. 

The commutators of $Q$ with the fermionic creation and annihilation 
operators are given by

\bea
\left[Q,b(\bfp,s)\right]&=&\beta_{ps}\;
e^{i\delta_{ps}}\;d^\dagger(-\bfp,s) \nonumber \\
\left[Q,d^\dagger(-\bfp,s)\right]&=&-\beta_{ps}\;e^{-i\delta_{ps}}\;b(\bfp,s)
\pkt \nonumber
\eea
which leads to the following relation between the transformed and the
original operators  
\bea\label{bdbemol}
 b(\bfp,s)&=&
\cos \beta_{\bfp,s} \; b_{\flat}(\bfp,s)+
\sin\beta_{\bfp,s}\; e^{i\delta_{\bfp,s}}\; d^\dagger_{\flat}(-\bfp,s)
\kma  \nonumber\\
 d^\dagger(-\bfp,s)&=&
-\sin\beta_{\bfp,s}\; e^{-i\delta_{\bfp,s}} \; b_{\flat}(\bfp,s)
+\cos\beta_{\bfp,s} \; d^\dagger_{\flat}(-\bfp,s)
\pkt
\eea
The operators $  b(\bfp,s), \;  d(\bfp,s) $ as well as $  b_{\flat}(\bfp,s),
 \;  d_{\flat}(\bfp,s) $ obey the usual canonical  anticommutation relations.

The initial density matrix is now given by
\be\label{mdens}
\rho_{\flat}(0) = e^{-Q}~ \rho(0) ~ e^{Q} =  |0_{\flat}><0_{\flat}|
\pkt\ee

Although we have focused on a pure (vacuum) state, obviously this can
be easily generalized to thermal or non-thermal mixed states. 

In Appendix A we provide the details that lead to the following Green's
functions in the Bogoliubov transformed states.

With this restriction the transformed Green function  becomes
\bea \nonumber
&&i
S^>_{\flat}(t,\bfx;t',\bfx')=\langle 0_{\flat}|\psi(t,\bfx)\bar\psi(t',\bfx') 
|0_{\flat}\rangle=  
\\ \nonumber
&&\intp e^{i\bfp(\bfx-\bfx')}
\left[\cos^2\!\beta_p \; (\slash\hspace{-2.3mm}p+m)e^{-i\Ep(t-t')}  
\right.
\\ \nonumber &&
-\sin\beta_p \; \cos\beta_p \;  e^{i\delta_p} \; 
\bSigma  \hat\bfp(\slash\hspace{-2.3mm}p+m)\gamma_5
\gamma_0
e^{-i\Ep(t+t')}\\ \nonumber &&
-\sin\beta_p \; \cos\beta_p \; e^{-i\delta_p} \; \bSigma \; \hat\bfp \; 
\gamma_5\gamma_0
(\slash\hspace{-2.3mm}p+m) \; 
e^{i\Ep(t+t')}\\ \nonumber &&
\left. +\; \sin^2\!\beta_p \; \gamma_5\gamma_0
(\slash\hspace{-2.3mm}p+m)\gamma_5 \gamma_0 \;  e^{i\Ep(t-t')}
\right] \label{Sgreatbogo}
\kma\eea

and 

\bea
&&-i S^<_{\flat}(t,\bfx;t',\bfx')=\langle
0_{\flat}|\bar\psi(t',\bfx')\psi(t,\bfx) 
| 0_{\flat} \rangle= 
\\ \nonumber
&&\intp e^{i\bfp(\bfx-\bfx')}\;
\left[\sin^2\!\beta_p \; (\slash\hspace{-2.3mm}p+m)\;e^{-i\Ep(t-t')}  
\right.
\\ \nonumber &&
\sin\beta_p\; \cos\beta_p\;e^{i\delta_p}\;\bSigma\hat\bfp
(\slash\hspace{-2.3mm}p+m)\gamma_5
\gamma_0\;
e^{-i\Ep(t+t')}\\ \nonumber &&
+\sin\beta_p\;\cos\beta_p\;e^{-i\delta_p}\;\bSigma\hat\bfp\;
\gamma_5\gamma_0
(\slash\hspace{-2.3mm}p+m)\;
e^{i\Ep(t+t')}\\ \nonumber &&
\left. +\cos^2\!\beta_p\;\gamma_5\gamma_0
(\slash\hspace{-2.3mm}p+m)\gamma_5 \gamma_0e^{i\Ep(t-t')}
\right] \label{Ssmallbogo}
\pkt
\eea

Perhaps the most striking feature of these Green's functions is their 
lack of time translational invariance, the main reason is that the
Bogoliubov transformed states are not eigenstates of the bare particle 
number. We note that in the terms with $ t+t'$ a translation of the time
variables can be compensated by a change in the phase $ \delta_p $, i.e. a
gauge transformation of the fermionic fields. 

 It is this lack of time translational invariance that will allow
to cancel the initial time singularities as shown explicitly below. 

The evolution equation is obtained in the same manner as in the previous
section and is  exactly of the same form as (\ref{evolequ}) with
${\bf q}=0$ as befits the homogeneous case, with the self-energy and
tadpole kernels now given by the expressions
(\ref{tadeq})-(\ref{selfenerequi}) but in terms of the Bogoliubov
transformed Green's functions 

\be
J_\flat (t)=-ig \tr  S^>_{\flat}(t,\bfx;t,\bfx)=- g\intp 
\left[ 4 m \cos2\beta_p
-(-4p)\sin2\beta_p\;\cos(2 \Ep t-\delta_p)\right]\label{tadpolebogo}
\pkt\ee
and
\bea\label{sigmab}
\Sigma_{\flat,\bf 0}(t,t')&=&-2 g^2\intp \frac{1}{2\Ep}
\left\{8\; p^2\cos(2\beta_p)\sin[2\Ep(t-t')]\right.
 \\ \nonumber
&&\left. -8\;p\;m\sin(2\beta_p)\left[\sin(2\Ep t -\delta_p)
-\sin(2\Ep t' -\delta_p)\right]\right\}
\pkt\eea

The first, time independent contribution to $ J_{\flat}(t) $ in
eq.(\ref{tadpolebogo}) can be absorbed into
a constant shift of the condensate much in the same way as in the
un-transformed case. The second, time dependent term will be
used to cancel the initial time singularities. 

The self-energy kernel (\ref{sigmab}) can be written as $\Sigma_{\bf
0}(t-t')+ \Delta \Sigma_{\flat,\bf 0}(t,t')$ with 
$\Sigma_{\bf 0}(t-t')$ given by (\ref{selfhomoeq}) and 

\bea
\Delta\Sigma_{\flat,\bf 0}(t,t')&=&-2g^2 \intp \frac{1}{2\Ep} 
\left\{8p^2\; (\cos2\beta_p -1)\sin[2\Ep(t-t')]\right.\nonumber
 \\ \nonumber
&&\left. -8\; p\;m\;\sin2\beta_p\left[\sin(2\Ep t -\delta_p)
-\sin(2\Ep t' -\delta_p)\right]\right\}\nonumber
\pkt\eea

We will assume in the following that $\beta_p$ decreases sufficiently
fast with $p$ so that terms proportional to $ \sin2\beta_p $ and
$ \cos(2\beta_p) -1 $ lead to convergent integrals. This in fact will
be checked a posteriori when we find the required expression for
$\beta_p$ below. Then the ultraviolet divergences in eq.(\ref{sigmab})
are the same as in the perturbative vacuum. 

Integrating by parts now three times in $ t' $ in order to single out
the ultraviolet and equal-time singularities from $ \Sigma_{\flat,\bf
0}(t,t') $ in the equation of motion, we find that just as in the
un-transformed case the ultraviolet divergences proportional to 
$ \phi_{\bf 0}(t) $ and $ \ddot{\phi}_{\bf 0}(t) $ are cancelled by the mass
and wave function renormalization counterterms (up to finite parts
depending on the renormalization prescription).

The terms that result in initial-time singularities arise from 
\bea
&& \left. \left[\int_0^t\; dt'\;
\Sigma_{\flat,\bf 0}(t-t')\;\phi_{\bf 0}(t')\right]\right|_{\rm sing} =
\nonumber\\ \nonumber
&&- g^2\intp\frac{8p^2}{\Ep} \left\{
-\frac{1}{2\Ep}\phi_{\bf 0}(0)\;\cos2\Ep t
-\frac{1}{(2\Ep)^2}\;\dot \phi_{\bf 0}(0)\;\sin2\Ep t \right .
\\ \nonumber && \left. 
+\frac{1}{(2\Ep)^3}\; \ddot \phi_{\bf 0}(0)\;
\cos2\Ep t \right\} \cos 2\beta_p
\eea 
We now require that these terms are cancelled by the time dependent
terms of  $J_\flat(t)$, eq.(\ref{tadpolebogo}). 
This requirement leads to the equation
\bea \nonumber
&&  \tan2\beta_p \; \cos(2\Ep t - \delta_p) =
\\ \nonumber
&& g\; \frac{2p}{\Ep} \left\{
-\frac{1}{2\Ep}\;\phi_{\bf 0}(0)\;\cos2\Ep t 
-\frac{1}{(2\Ep)^2}\;\dot \phi_{\bf 0}(0)\;\sin2\Ep t \right.
\\ \nonumber && \left. 
+\frac{1}{(2\Ep)^3}\; \ddot \phi_{\bf 0}(0)\;
\cos2\Ep t
\right\} 
\eea

\noindent Comparing the terms proportional to the sine and cosine we
find two equations that determine 
 $\delta_p ~,~\beta_p$. These equations can be solved perturbatively
with 
$$
\beta_p = b_{1,p}\; g+ b_{2,p}\; g^2+ {\cal O}(g^3)\;\;\; .
$$
To one loop order we only need to keep the linear term in $g$  leading to

\bea \nonumber
\beta_p \cos\delta_p &=& g\; \frac{p}{\Ep} \left[
-\frac{1}{2\Ep}\;\phi_{\bf 0}(0)
+\frac{1}{(2\Ep)^3}\; \ddot \phi_{\bf 0}(0)\;
 \right] \\
\beta_p \sin\delta_p &=& g\;\frac{p}{\Ep} \left[
-\frac{1}{(2\Ep)^2}\;\dot \phi_{\bf 0}(0) \right] \label{bogocoeffshomo}
\eea

At this stage we recognize that there is freedom in choosing the
Bogoliubov parameters to cancel the initial time singularities. The
initial value problem will be free of singularities by choosing the
coefficients $ \beta_p \; ; \; \delta_p $ so as to cancel the terms
proportional to $ 1/p ~, 1/p^3 $ and so that $ \beta_p $ vanishes as $
p \rightarrow \infty$.

Different  choices of the Bogoliubov parameters that
differ only in higher inverse  powers of the momenta  lead to
different initial quantum states, but the initial value problem is
free of initial singularities. This freedom is similar to choosing
renormalization counterterms including finite parts, i.e. different
renormalization prescriptions.

Now the Bogoliubov correction and the mass and wave-function renormalization 
counter terms remove exactly all ultraviolet and initial time
divergences from the equation of motion. 
To lowest order we can set $\beta_p=0$ in the self-energy in the
equation of motion and having removed all ultraviolet 
and initial time singularities and absorbing the time independent
contribution from the tadpole term in a constant 
shift of the condensate, we finally obtain the evolution equation in
the case of the homogeneous condensate 
 
\be \label{reneqm}
\ddot \phi_{\bf 0}(t) + M^2\; \phi_{\bf 0} (t) 
+ \int_0^t dt'\;\Sigma_s (t-t')\stackrel{\dots}{\phi}_{\bf 0}(t')=0
\ee
where  $\Sigma_s(t-t') $ is the subtracted self energy kernel
\bea 
\Sigma_s(t-t') =  - g^2\intp\frac{p^2}{\Ep^4}
\; \cos\left[2\Ep(t-t')\right]\label{bogofinker}
\pkt \eea

The equation of motion above should be compared to Eqs (7.21) and (7.22)
in \cite{Baacke:1998c}. The equation of motion obtained there
differs from (\ref{reneqm}) by a {\em finite} renormalization $\Delta Z$ 
that corresponds to a different renormalization scheme but otherwise
the equations are the same up to one loop order.  

We note that even in the massless limit the tadpole
$J_{\flat}=\langle\bar\psi \psi\rangle$ 
is nonvanishing, signaling chiral symmetry breaking, which in this
case  is solely a consequence of the initial conditions. This is an 
important result of our analysis, that non-equilibrium initial states can lead
to symmetry breaking.

\section{Linear response theory}
\setcounter{equation}{0}

The initial value problem can be obtained by establishing direct
contact with linear response theory as presented in 
ref.\cite{boyRG} in the case of scalar condensates and in ref.\cite{shang}
for fermionic coherent states. This is achieved by coupling an {\em
external source} term to the scalar field in the Lagrangian density
(\ref{lagra}) 
$$
{\cal L}[\Phi,\psi,\bar\psi] \rightarrow {\cal
L}[\Phi,\psi,\bar\psi]+J_{ext}(x) \; \Phi(x) \; .
$$
The expectation value of the scalar field induced by this source term
is given by 

\be
\langle \Phi(x) \rangle = i\int dx' J_{ext}(x')\left[\langle
\Phi^+(x)\Phi^+(x') \rangle - \langle \Phi^+(x)\Phi^-(x')\rangle
\right] \label{linresp} 
\ee
\noindent where the superscripts $\pm$ refer to the forward and
backward time branches in the real time generating 
functional. The bracket in (\ref{linresp}) is the retarded
commutator. As discussed in detail in ref.\cite{boyRG} the inversion
of (\ref{linresp}) gives rise to the equation of motion for the scalar
field with an inhomogeneity given by the external source
term. Following the same steps detailed in the first section, we find
the equation of motion for an homogenous condensate to be given by

\bea \nonumber
(1+\delta Z )\; \ddot\phi_{\bf 0}(t)
+(M^2+\delta M^2)\; \phi_{\bf 0}(t)+\int_{-\infty}^t dt'\;\Sigma_{\bf
0}(t-t')\;\phi_{\bf 0}(t')-J =J_{ext,\bf 0}(t)
\kma\eea
which differs from (\ref{evolequ}) in the lower limit in the non-local
term with the self-energy. Assuming adiabatic 
switching-on of the interaction from $t=-\infty$ we can now rewrite
this equation of motion in a form that is closer to (\ref{evolequ}) by
integrating by parts the non-local term. Definining, 

\bea\nonumber
\Sigma_{\bf 0}(t-t')= \frac{d}{dt'}\Sigma_{1,\bf 0}(t-t')=
\frac{d^2}{dt'^2}\Sigma_{2,\bf 0}(t-t')=
\frac{d^3}{dt'^3}\Sigma_{3,\bf 0}(t-t')
\eea
and integrating by parts we obtain
\bea 
&&(1+\delta Z )\;\ddot\phi_{\bf 0}(t)
+(M^2+\delta M^2)\;\phi_{\bf 0}(t)
+\Sigma_{1,\bf 0}(0)\;\phi_{\bf 0}(t)
-\Sigma_{2,\bf 0}(0)\;\dot\phi_{\bf 0}(t)
+\Sigma_{3, \bf 0}(0)\;\ddot\phi_{\bf 0}(t)\nonumber\\ 
&&+\int_{-\infty}^t dt'\;
\Sigma_{s,\bf 0}(t-t')\stackrel{\dots}{\phi}_{\bf 0}(t')-J= J_{ext,\bf
0}(t)\; 
\kma\nonumber\eea
with $\Sigma_{s,\bf 0}(t-t')$ given by (\ref{bogofinker}).
Using the explicit form of $ \Sigma_{\bf 0}(t) $ given by
(\ref{selfhomoeq}) we find  
\bea 
\Sigma_{1,\bf 0}(0)&=&- g^2\int\frac{d^3p}{(2\pi)^3 2\Ep}\frac{4 p^2}{\Ep^2}
=-\delta M^2 \label{masscount}
\\
\Sigma_{2,\bf 0}(0)&=&0
\nonumber\\ 
\Sigma_{3,\bf 0}(0)&=&g^2\int\frac{d^3p}{(2\pi)^3 2\Ep}\frac{4 p^2}{\Ep^4}
=-\delta Z \label{wavefuncount}
\eea

Therefore the specific choice of external current
\be\label{corribo}
J_{ext}(t)=\int_{-\infty}^{0}dt'\; \Sigma_s (t-t')\;
\stackrel{\dots}{\phi}_{\bf 0}(t') \label{timesource} 
\ee
leads to the initial value problem described by eq.(\ref{reneqm}). 

This current depends on the past history of the condensate, and in general
{\em does not vanish for $ t>0 $} as it would be desirable from the point
of view of linear response. In linear response the initial value problem
is envisaged to be prepared by switching-on an external  source and
the interaction adiabatically from $t=-\infty$. The external source
acts as a Lagrange multiplier, slowly displacing the condensate to the
value to be determined at $ t=0 $, and 
switching-off the source suddenly at this time.  This allows the condensate
to be formed and dressed over a very long period of time. The dressed
condensate is then released when the external current is switched-off.
However, in an interacting renormalizable theory this instantaneous
switching-off of the external current results in singularities. To set
up a consistent, singularity free and renormalized initial value problem as
is the goal of this article, the choice of the current (\ref{corribo}) is the
one that establishes contact with a linear response formulation, in this
case the current depends on the past history of the condensate, which
obviously need not be specified for an initial value problem. On the 
other hand, such a choice of current, depending on the past history is
rather artificial from  the linear response point of view.  

We note that the current can be made to vanish at all times with the 
particular choice
\bea \nonumber
\phi_{\bf 0}(t)=\phi_{\bf 0}(0)+\dot\phi_{\bf 0}(0)\;t+\frac12 \,
\ddot\phi_{\bf 0}(0)\;t^2  
\eea
\noindent for  $t<0$. Obviously this behaviour manifests   
 the problem of the initial time singularities  as singularities in
the behavior of the field at a  remote past. 

An alternative would be to assume that the external source and
therefore the condensate is adiabatically switched-on with a damping
factor $ e^{\epsilon t} $ for $ t<0 $ but this results in {\em
discontinuities} in the first or second derivatives at $ t=0 $ and
that would produce additional contributions from the integration by
parts from these discontinuities.  

There are two main conclusions of this discussion on the relationship
with linear response:
\begin{itemize}
\item{We have established a {\bf direct} relationship between the
evolution equations in the linearized approximation and linear
response theory. The initial value problem free of UV and {\bf initial time}
singularities is shown to be obtained in the context of linear
response through a particular choice of the external current.
Such  external current depends on the 
past history of the condensate and only a very specific form for 
it leads to a singularity free initial value problem.
From the perspective of an initial value 
problem in which Cauchy data are specified at some given initial time
on a space-like hypersurface this is a consistent choice. However,  from the
point of view of linear response this choice is somewhat
artificial. The resulting external current {\em does not vanish} for $
t>0 $. If we instead require  an instantaneous switching-off of the
external current at $ t=0 $,   initial time singularities become unavoidable.}

\item{The method of preparing a dressed initial state via a Bogoliubov
transformation leads to a satisfactory description of the initial
value  
problem. The usual mass, wave function and coupling constant
renormalization counterterms cancel the ultraviolet divergences, and
the Bogoliubov coefficients are judiciously chosen to cancel the
initial time divergences consistently in perturbation theory. To
lowest order in the Yukawa coupling and for a homogenous condensate
such a choice is given by (\ref{bogocoeffshomo}).} 
\end{itemize} 

Having studied in detail the simpler case of the homogeneous condensate,
we now move on to our main point, the study of the evolution of
inhomogeneous condensates.  

\section{Equations of motion for inhomogenous condensates}
\label{linres_inh}
\setcounter{equation}{0}

The  equation of motion for non-homogeneous condensates in the
amplitude approximation and in terms of spatial Fourier transforms reads
\bea\nonumber
&&(1+\delta Z )\left[\ddot \phi_{\bf q}(t)+{\bf q}^2\; 
 \phi_{\bf q}(t)\right] + (M^2 +\delta M^2)\;\phi_{\bf q} (t) 
\\ \nonumber&& + \int_0^t dt'\; \Sigma_{\bf q} (t,t')\; 
 \phi_{\bf q}(t') +J_{\flat,\bf q}(t)=0
\pkt\eea

From the discussions of the previous sections we have learned that
the Bogoliubov coefficients that define the dressed states at the initial
time can be found consistently in perturbation theory. Since the self-energy
is already of second order in the Yukawa coupling we will not need to
consider the Bogoliubov corrections to the self-energy, but only to the
tadpole term $J_{\flat,\bf q}(t)$.

The one-loop self-energy is therefore the usual one and given by
\bea 
i\Sigma_{\bf q}(t-t') &=& 4
 g^2 \int \frac{d^3p}{(2\pi)^3 2E_{\bf p}2E_{{\bf p}-{\bfq}}}
 \left[ E_{\bf p}+E_{{\bf p}-{\bfq}}+\bfp({\bf p}-{\bf q})-m^2\right]
\nonumber \\  &&
(-2i)\sin\left[(E_{\bf p} + E_{{\bf p}-{\bfq}})(t-t')\right]
\pkt\eea

The derivation of the tadpole diagram that determines $J_{\flat,\bf
q}(t)$ using the Bogoliubov-transformed Green functions, is given in
Appendix B. We find 
\bea\nonumber 
&&J_{\flat,{\bf q}}(t)= 
4 \; m \; g\intp  
\\ \nonumber &&-2 \int \frac{d^3p}{(2\pi)^3 2E_{\bf p}2E_{{\bf p}-{\bfq}}}
 \left[
\gamma^*({\bf q},\bfp) \; 
e^{-i(E_{\bf p} + E_{{\bf p}-{\bfq}})t} +\gamma({\bf q},\bfp')  \; 
e^{i(E_{\bf p} + E_{{\bf p}-{\bfq}}) t}\right]
\pkt \eea 
The function  $\gamma({\bf q},\bfp)$ is a function related
to the angles of the Bogoliubov transformation, that will be specified below 
such as to remove the initial time singularities.  

The analysis of the singular and divergent contributions
in the equation of motion
proceeds as in the homogenous case,  performing three integrations
by parts with respect to the time in the non-local term, we find
\bea\nonumber 
&& \int_0^t dt' \;  \sin[(E_{\bf p} + E_{{\bf p}-{\bfq}}) (t-t')] \;
\phi_{\bf q}(t') = 
\frac{\phi_{\bf q}(t)}{E_{\bf p} + E_{{\bf p}-{\bfq}}}
-\frac{\phi_{\bf q}(0)}{E_{\bf p} + E_{{\bf p}-{\bfq}}}
\cos[(E_{\bf p} + E_{{\bf p}-{\bfq}}) t] 
 \nonumber \\\nonumber 
&&-\frac{\dot \phi_{\bf q}(0)}{(E_{\bf p} + E_{{\bf p}-{\bfq}})^2}
\sin[(E_{\bf p} + E_{{\bf p}-{\bfq}}) t]
 -\frac{\ddot \phi_{\bf q}(t)}{(E_{\bf p} + E_{{\bf p}-{\bfq}})^3}
\\ 
&&+\frac{\ddot \phi_{\bf q}(0)}{(E_{\bf p} + E_{{\bf p}-{\bfq}})^3} 
\cos[(E_{\bf p} + E_{{\bf p}-{\bfq}}) t]+\frac{1}{(E_{\bf p} + E_{{\bf
p}-{\bfq}})^3} 
\int_0^t dt' \cos[(E_{\bf p} + E_{{\bf p}-{\bfq}})(t-t')] \; 
\stackrel{\dots}{\phi_{\bf q}}(t')\nonumber
\pkt\eea
The parts containing $\phi_{\bf q}(0)$ and its derivatives lead to
initial time singularities in the equation of motion, these can be
isolated by writing 
\bea\nonumber 
&&\left[ \int_0^t dt'\Sigma_{\bf q} (t-t')
 \phi_{\bf q}(t')\right]_{\rm sing}= \\ \nonumber 
&&-8 g^2 \int \frac{d^3p}{(2\pi)^3 2E_{\bf p}2E_{{\bf p}-{\bfq}}}
\left[E_{\bf p}  E_{{\bf p}-{\bfq}}+{\bf p}({\bf p}-{\bf
q})-m^2\right]  \\ \nonumber
&&\left[
\tau({\bf q},\bfp)  \; 
e^{-i(E_{\bf p} + E_{{\bf p}-{\bfq}})t} +\tau^*({\bf q},\bfp)  \; 
e^{i(E_{\bf p} + E_{{\bf p}-{\bfq}}) t}\right]
\eea
with
\bea\nonumber 
\tau({\bf q},\bfp)= \frac{1}{2}\left[
-\frac{\phi_{\bf q}(0)}{E_{\bf p} + E_{{\bf p}-{\bfq}}} 
+i\frac{\dot \phi_{\bf q}(0)}{(E_{\bf p} + E_{{\bf p}-{\bfq}})^2}
+\frac{\ddot \phi_{\bf q}(0)}{(E_{\bf p} + E_{{\bf p}-{\bfq}})^3} \right]
\; ;\eea
note that $ \phi_{\bf q}(0)=\phi^*_{\bf -q}(0) $.  
With the choice
\bea\nonumber 
\gamma({\bf q},\bfp)=-4g\left[
E_{\bf p}+E_{{\bf p}-{\bfq}}+\bfp({\bf p}-{\bf q})-m^2\right]
\tau^*({\bf q},\bfp)
\eea
the current $J_{\flat,{\bf q}}(t)$  exactly  cancels the initial time
singularities in the non-local term with the self-energy.
As in the homogeneous case, the current $ J_{\flat,\bf q}(t) $, which is a
Fourier transform of $ \langle \psi(\bfx)\psi(\bfx')\rangle $ is nonvanishing. 
Again,  chiral symmetry is here broken by the initial conditions.

The ultraviolet divergent contributions of the self energy to
 the equation of motion are given by
\bea\nonumber 
&&\left[ \int_0^t dt' \; \Sigma_{\bf q} (t-t') \;
 \phi_{\bf q}(t')\right]_{\rm UV~div}= \\ \nonumber 
&&-8 g^2\int \frac{d^3p}{(2\pi)^3 2E_{\bf p}2E_{{\bf p}-{\bfq}}}
 \left[E_{\bf p}+E_{{\bf p}-{\bfq}}+\bfp({\bf p}-{\bf q})-m^2\right]
\\ \nonumber
&&\left[\frac{\phi_{\bf q}(t)}{E_{\bf p}+E_{{\bf p}-{\bfq}}}-
\frac{\ddot{\phi}_{\bf q}(t)}{(E_{\bf p}+E_{{\bf p}-{\bfq}})^3}
\right]\\ \nonumber
&&=
\tilde\Sigma_1(\bfq^2) \; \phi_{\bf q}(t)+
\tilde\Sigma_3(\bfq^2) \; \ddot{\phi}_{\bf q}(t)
\pkt\eea
Here we  define the UV divergent parts of the self energy kernel in
dimensional regularization as
\bea\nonumber 
\tilde\Sigma_1(\bfq^2)&=&-8 \;g^2
\int\frac{d^{3-\epsilon}p}{(2\pi)^{3-\epsilon}}
\frac{E_{\bf p}+E_{{\bf p}-{\bfq}}+\bfp({\bf p}-{\bf q})-m^2}{4E_{\bf
 p}E_{{\bf p}-{\bfq}}(E_{\bf p}+E_{{\bf p}-{\bfq}})}\\\nonumber  
\tilde\Sigma_3(\bfq^2)&=&8 \;g^2\int\frac{d^{3-\epsilon}p}
{(2\pi)^{3-\epsilon}}
\frac{E_{\bf p}+E_{{\bf p}-{\bfq}}+\bfp({\bf p}-{\bf q})-m^2}{4E_{\bf
 p}E_{{\bf p}-{\bfq}}(E_{\bf p}+E_{{\bf p}-{\bfq}})^3} 
\pkt\eea 
These expressions have to be regularized to obtain the
renormalized equation of motion.
This is discussed in detail in Appendix C.
The singular and ultraviolet divergent parts are cancelled by the
appropriate choice of the mass and wave function renormalization and the
Bogoliubov coefficient in the tadpole. The final form of the subtracted
 self-energy kernel is given by
\bea\nonumber 
\Sigma_{s,{\bf q}}(t-t')= &&
-8 g^2\int \frac{d^3p}{(2\pi)^3 2E_{\bf p}2E_{{\bf p}-{\bfq}}}
\\ \nonumber &&
\frac{E_{\bf p}+E_{{\bf p}-{\bfq}}+\bfp({\bf p}-{\bf q})-m^2}{(E_{\bf
 p}+E_{{\bf p}-{\bfq}})^3} \;  
\cos[(E_{\bf p}+E_{{\bf p}-{\bfq}})(t-t')]
\kma\eea
\noindent It follows from rotation invariance that the 
kernel $ \Sigma_{s,{\bf q}}(t) $ only depends on $ \bf q^2 $.

The equation of motion in momentum space becomes
\bea\nonumber
&&[1+\delta Z+\tilde \Sigma_3( {\bf q}^2)]\ddot\phi_{\bf q}(t)
+\left[q^2(1+\delta Z)+
M^2+\delta M^2 +\tilde\Sigma_1( {\bf q}^2)\right]
\phi_{\bf q}(t) \\ \nonumber
&&+\int_0^t dt'\; \tilde\Sigma_{s,\bf q}(t-t')\dddot\phi_{\bf q}(t')
=0
\pkt\eea 
We show in Appendix C that
we can decompose $ \tilde\Sigma_1({\bf q}^2) $ and $
\tilde\Sigma_3({\bf q}^2) $ as 
\bea
\Sigmat_1({\bf q}^2)&=&-\delta M^2 - {\bf q}^2 \; \delta Z
+\Delta \Sigmat_1({\bf q}^2) \nonumber \\
\Sigmat_3({\bf q}^2)&=&-\delta Z +\Delta\Sigmat^3({\bf q}^2)\nonumber
\eea
The divergent and finite parts are explicitly given in Appendix C.
We finally obtain the renormalized equation of motion which is
free from ultraviolet and initial time singularities

\bea \nonumber 
&&[1+\Delta\Sigmat_3({\bf q}^2)] \;\ddot\phi_{\bf q}(t)
+\left[\bfq^2+
M^2+\Delta\Sigmat_1({\bf q}^2)\right]
\phi_{\bf q}(t) \kma \\ 
&&+\int_0^t dt'\; \Sigmat_{s,{\bf q}}(t-t')\dddot\phi_{\bf q}(t')
=0 \label{eqm_inh_ren}
\kma \eea
where
\begin{equation}
\Sigmat_{s,{\bf q}}(t-t') =-8 \; g^2\int \frac{d^3p}{(2\pi)^34E_{\bf
p}E_{{\bf p}-{\bf q}}} 
\frac{E_{\bf p}E_{{\bf p}-{\bf q}}+\bfp(\bfp-{\bf q})-m^2}{(E_{\bf
p}+E_{{\bf p}-{\bf q}})^3} 
\cos\left[(\Ep+\Epp)\tau\right] \label{sigmasubinho}
\pkt\end{equation}
\section{Solution of the equation of motion: numerical analysis}
\label{solu}
\setcounter{equation}{0}
We have derived in the previous section the renormalized equation of
motion. It can be solved in a standard way via Laplace transform. We introduce
\bea\nonumber 
\tilde\psi_{\bf q}(s)=\int_0^\infty dt\; e^{-st}\;\phi_{\bf q}(t)
\pkt\eea
for the condensate, and
\bea\nonumber 
\tilde \sigma_s(s^2,q^2)=s \int_0^\infty dt\; e^{-st}\;\tilde
\Sigma_{s,q}(t)
\eea
for the self energy\footnote{The extra factor $s$ is introduced so as to
make $\tilde \sigma_s$ a function of $s^2$.}. We find, 
\bea
\sigmat_s(s^2,\bfq^2)&=&-8g^2s
\int_0^\infty d\tau\; e^{-s\tau}
\int \frac{d^3p}{(2\pi)^34E_{\bf p}E_{{\bf p}-{\bf q}}}
\frac{E_{\bf p}E_{{\bf p}-{\bf q}}+\bfp(\bfp-{\bf q})-m^2}
{(E_{\bf p}+E_{{\bf p}-{\bf q}})^3}
\cos\left[(E_{\bf p}+E_{{\bf p}-{\bf q}})\tau\right]
\nonumber \\ \label{self_lap}
&=&-8g^2 s^2 \int \frac{d^3p}{(2\pi)^34E_{\bf p}E_{{\bf p}-{\bf q}}}
\frac{E_{\bf p}E_{{\bf p}-{\bf q}} +\bfp(\bfp-{\bf q})-m_2}
{(E_{\bf p}+E_{{\bf p}-{\bf q}})^3
[(E_{\bf p}+E_{{\bf p}-{\bf q}})^2+s^2]}
\pkt
\eea
The Laplace transformed renormalized equation of motion becomes
\bea \nonumber 
\left\{s^2\left[1+\Delta\tilde\Sigma_3(\bfq^2)
+\sigmat_s(s^2,\bfq^2)\right]+
\bfq^2+M^2+\Delta\tilde\Sigma_1(\bfq^2)\right\}
\tilde\psi_{\bf q}(s)
\\ \nonumber
=\left[\dot\phi_{\bf q}(0)+s\phi_{\bf q}(0)\right]
\left[1+\Delta\tilde\Sigma_3(\bfq^2)
+\sigmat_s(s^2,\bfq^2)\right]
+\frac{\ddot\phi_{\bf q}(0)}{s}\; \sigmat_s(s^2,\bfq^2)
\kma\eea
so that
\bea\nonumber
\tilde\psi_{\bf q}(s)=\frac{\left[\dot\phi_{\bf q}(0)+s\phi_{\bf q}(0)
\right]\left[1+\Delta\Sigmat_3(\bfq^2)+
\sigmat_s(s^2,\bfq)\right]
+\ddot\phit(0,\bfq)\; \sigmat_s(s^2,\bfq^2)/s}
{s^2\left[1+\Delta\Sigmat_3(\bfq^2)
+\sigmat_s(s^2,\bfq^2) \right]+
\bfq^2+M^2+\Delta\Sigmat_1(\bfq^2)}
\pkt\eea
The solution $\phi_{\bf q}(t)$ then is
obtained by the inverse transformation
\bea\nonumber
\phi_{\bf q}(t)
=\int_{-i\infty+c}^{i\infty+c}\frac{ds}{2\pi i}\;
e^{st}\; \tilde\psi_{\bf q}(s)
\pkt\eea
This solution is discussed in detail in Appendix D. The integral above
is along the Bromwich contour with $c$ a positive real constant to the
right of all the singularities of the Laplace transform.  The  result
can be written as  (see Appendix D for details).
\bea\nonumber
\phi_{\bf q}(t)&=&\frac{2}{\pi}\int_{0^+}^\infty d\omega
\left[\cos(\omega t) \; \phi_{\bf q}(0) \;\omega \; \im
F_1(-\omega^2+io,\bfq^2) \right.\\ \nonumber
&&+\sin(\omega t) \; \dot \phi_{\bf q}(0) \; \im F_1(-\omega^2+io,\bfq^2)
\\ \nonumber
&&\left.-\frac{\ddot\phi_{\bf q}(0)}{\omega} \; \cos(\omega t) \;
\im F_2(-\omega^2+io,\bfq^2)\right]\kma
\eea
where
\bea\nonumber
F_1(s^2,\bfq^2)&=&\frac{1+\Delta\Sigmat_3(\bfq^2)+
\sigmat_s(s^2,\bfq^2)}
{s^2\left[1+\Delta\Sigmat_3(\bfq^2)+
\sigmat_s(s^2,\bfq^2)\right]+\bfq^2+M^2+\Delta\Sigmat_1
(s^2,\bfq^2)}
\\\nonumber
F_2(s^2,\bfq^2)&=&\frac{\sigmat_s(s^2,\bfq^2)}
{s^2\left[1+\Delta\Sigmat_3(\bfq^2)+
\sigmat_s(s^2,\bfq^2)\right]+\bfq^2+M^2+\Delta\Sigmat_1
(\bfq^2)}  
\pkt\eea
$ \ddot\phi_{\bf q}(0) $ is not an independent initial 
value, it is determined by setting $ t=0 $ in the equation of motion
(\ref{eqm_inh_ren}) with the result
\bea\nonumber
\ddot\phi_{\bf q}(0)=-\phi_{\bf q}(0)\;\frac{\bfq^2+M^2+
\Delta\Sigmat_1(\bfq^2)}{1+\Delta\Sigmat_3(\bfq)^2}
\pkt\eea
It is then convenient to define a kernel $F_3$ which
combines $F_1$ and $F_2$ with prefactors uniqueley determined
by $\phi_{\bf q}(0)$. With the definition
\bea\nonumber
F_3(\omega,\bfq^2)=-\frac{1}{\omega}+
\frac{\bfq^2+M^2+\Delta\Sigma_1}{\left(
 1+\Delta\Sigma_3\right)\omega} \times\frac{1}{-\omega^2+\frac{
\dss \bfq^2+M^2+\Delta\Sigma_1}
{\dss 1+\Delta\Sigma_3+\sigmat_s(-\omega^2+io,\bfq^2)}}
\eea
the solution can be written as
\bea\nonumber
\phi_{\bf q}(t)=\frac{2}{\pi}\int_{\omega_c}^\infty d\omega
\left[\phi_{\bf q}(0)\cos(\omega t) \;  \im F_3(\omega-io,\bfq^2)+\dot
\phi_{\bf q}(0) \; \sin(\omega t) \; \im F_1(-\omega^2+io,\bfq^2) \right]
\eea

The function $F_3$ which is, up to prefactors, the Laplace 
transform of the solution, exhibits a pole at
\bea\nonumber
\omega_R = \omega_{\bf q}+ \delta \omega ~~;~~ \omega_{\bf q} =
\sqrt{M^2+\bfq^2} 
\kma\eea 
with $ \delta \omega = {\cal O}(g^2) $. If $M > 2 m$  the scalar field
can decay into a fermion-antifermion pair, the pole actually describes
a resonance. In perturbation theory the width of this resonance
is perturbatively small and near the resonance we can approximate 
the function $F_3$ by a Breit-Wigner resonance

\bea\nonumber
F_3(\omega,\bfq^2)\simeq\frac{1}{2}
\frac{Z_{\rm R}}{\omega-\omega_{\rm R}
-i \; \Gamma_{\rm R}  }
\pkt\eea
The resonance position is determined by
\bea\nonumber
\omega_{\rm R}=\re \frac{\omega^2_{\bf q}+\Delta\Sigma_1}
{1+\Delta\Sigma_3+\sigmat_s(-\omega^2_{\bf q}+io,\bfq^2)}
\kma\eea
the residue $Z_{\rm R}$ is given by
\bea\nonumber
Z_{\rm R}=
\frac{\omega^2_{\bf q}+\Delta\Sigma_1}
{\left(1+\Delta\Sigma_3\right)\omega_{\bf q}} \frac{1}{-2\omega_{\bf q}+
\frac{\dss d}{\dss d\omega} \left[\re\frac{\dss \omega^2_{\bf
q}+\Delta\Sigma_1} { \dss 1+\Delta\Sigma_3+\sigmat_s(-\omega^2+io,\bfq^2)}
\right]_{\omega=\omega_{\bf q}}}
\eea
and the width by
\bea\nonumber
\Gamma_{\rm R}=\frac{1}{2\omega_{\rm R}}\im \left[\frac{\omega^2_{\bf
q}+\Delta\Sigma_1} {1+\Delta\Sigma_3+
\sigmat_s(-\omega^2+io,\bfq^2)}\right]_{\omega=\omega_{\bf q}}
\pkt\eea

{\bf Numerical Analysis: }

\bigskip

We are now in conditions to study the evolution of an initial scalar 
condensate numerically by performing the
inverse Laplace and Fourier transforms since all the quantities are 
given by the subtracted one-loop self-energy. 

We consider two separate cases: $ M>2m $ in which case the scalar 
can decay into fermion-antifermion pairs, and $ M<2m $
in which case the scalar particle is stable. In both cases we studied 
the evolution for an initial spherical wave of gaussian profile 
\bea
\phi(0,\bfx)=N_0 \exp(-\bfx^2/2R_0^2) ~~; ~~ 
\dot\phi(0,\bfx)=0 ~~;~~ \mbox{with} ~~\int d^3x \; \phi(0,\bfx)=1 
\label{gaussian}
\eea

Since this gaussian wave-packet has zero center of mass momentum, 
the peak of the wave packet will not displace under
time evolution, but the wave packet will disperse and spread 
out in space-time. 

{\bf $ M>2m $:} We have chosen $ M=3m $ but there is a rather smooth 
variation of the wave function renormalization constant,
position and width of the resonance for  reasonable values of 
the ratio $ 2< M/m\leq 10 $. As a first step we calculated  the value
of the wave function renormalization numerically and find that 
 $ Z_{\rm R} $ differs from unity by less than
$ 3\% $ and that the ratio of the width to the position of 
the pole $\Gamma/\omega_{\rm R} \approx 0.02 $ for $ g =1 $ or smaller
and $ 2< M/m \leq 10 $. Therefore if the coupling is of this order or smaller,
the  approximation of the spectral density (discontinuity across the
cut) by the imaginary part of the Breit-Wigner form given above is 
excellent. The agreement obtained from the full numerical evolution
and that obtained from the approximate Breit-Wigner form is excellent. 
The evolution in this case is depicted in Fig. 1 that
displays the profile of the condensate as a function of $ t $ 
and $ |\vec x| $ for $ M=3; m=1; g=1 $ . We see that the propagation 
is inside the light cone and damped in time.   The peak at the origin 
is the center of mass of the wave packet, it oscillates in time with 
a frequency $ \approx \omega_R $ and decays on a time scale $
\Gamma^{-1} $. The evolution at very early times is smooth.

{\bf $ M<2m $:} In this case the scalar field is stable, 
the spectral density features  poles
 at $\omega=\pm \omega_{\rm R}$ below the fermion-antifermion cut.   
We find again that
$Z_{\rm P} \approx 0.97$, implying, that the contribution from the 
fermion-antifermion continuum is negligible except for small $t$.
Fig. 2 displays the time evolution of the gaussian wave packet in this case. 
The wave packet spreads in space-time, the evolution is always below
the light cone as clearly illustrated in the figure. The amplitude of
the wave packet decreases as a result
of spreading. Its spatial integral, which is equal to $\phi_{\bf 0}(t)$,
asymptotically oscillates with the pole frequency and amplitude $Z_P$. 

\section{Scalar Theories}
The connection between the  preparation of the initial state
via Bogoliubov transformations and the formulation in terms of 
linear response allows to generalize the study presented above to 
scalar theories. In particular let us consider the case of a scalar
 self-coupled  $\lambda \Phi^4 $ model using the linear response analysis. 
Focusing on the evolution equation for a  homogeneous condensate 
to order $\lambda^2$ we find 

\begin{eqnarray}\nonumber
&&(1+\delta Z)\; \ddot{\phi}_{\bf 0}(t)+(M^2+\delta M^2)\;
\phi_{\bf 0}(t)+\Sigma_{1,{\bf 0}}(0)\;\phi_{\bf 0}(t)
+\Sigma_{3,{\bf 0}}(0)\;\ddot{\phi}_{\bf 0}(t) \nonumber \\
&+&\int_{-\infty}^t dt'\; \Sigma_{s,{\bf 0}}(t-t')
\stackrel{\cdots}{\phi}_{\bf 0}(t')= J_{ext,{\bf 0}}(t)\;.  \label{fi4linres}
\end{eqnarray}   
The order $\lambda$ tadpole has been absorbed into $\delta M^2$ as 
a mass renormalization and the self energy to order $\lambda^2$ is 
given by the sunset diagram. The choice of counterterms 
$\delta Z, \delta M^2$ are as given in equations 
(\ref{masscount})-(\ref{wavefuncount}) and the external source 
introduced for linear response is specifically chosen as in 
eq.(\ref{timesource}). In this manner we obtain the initial 
value problem free of ultraviolet and initial time singularities. 
An alternative interpretation of the external source associated 
with the linear response formulation can be provided by 
introducing a Hartree factorization in the Lagrangian with a 
term of the form $ 4 \; \lambda \; \Phi \; \langle \Phi^3 \rangle $. 
This term acts now as an explicit source in the linearized 
equation of motion which is chosen so as to lead to a singularity 
free initial value problem, thus requiring a non-trivial 
Bogoliubov vacuum. The linear response analysis leads very 
simply to a well defined initial value problem for a proper choice of
the external source. The equivalence between the preparation of the
state via a Bogoliubov transformation of the free field Fock vacuum
(or density matrix) and the initial value problem obtained from 
linear response for the proper choice of external source allows now to
generalize the results obtained above for the linearized approximation. 

\section{Conclusions}

The non-equilibrium evolution of condensates in real time requires 
to provide Cauchy data at some initial time, hence
an initial value problem which requires the specification of an 
initially prepared quantum state or density matrix. An initial pure 
or mixed state of free field Fock quanta leads to initial time 
singularities. These have a simple interpretation: the evolution  
of expectation values or matrix elements in the interacting 
theory implies that the interaction is switched-on suddenly. The
interaction rearranges the spectral densities of the fields and the response to
the sudden switching-on of the interaction results in initial  
time singularities which are enhanced in a renormalizable theory. For
ystems with a finite number of degrees of freedom such effects are also
present but no singularities arise.

In summary, the time evolution of  {\bf arbitrary} quantum states or
density matrices in (interacting) field theory leads to {\bf short time
divergences}. Only for appropriately prepared pure or mixed initial
states, as those considered in this paper, the time evolution is well defined.
By appropriately prepared we mean states where  the filling for high
energy quanta follows a precise law determined by the initial data,
couplings and masses. 

We have chosen to study these initial time singularities 
and provide a consistent resolution in a Yukawa theory in $3+1$
dimensions, this theory being renormalizable allows to identify 
all of the divergences and singularities: ultraviolet
divergences associated with mass, coupling and wave-function 
renormalizations and initial time singularities that cannot
be cancelled by the usual counterterms. 

After recognizing the initial time singularities and their physical 
significance in the case of homogeneous condensates,
we have proposed a rather simple approach to provide a singularity 
free initial value problem. We introduced Bogoliubov
transformed initial states that incorporate the effects 
of dressing of states by the interaction. The Bogoliubov
coefficients can be obtained in a systematic series expansion 
in the Yukawa coupling and we have obtained them to one-loop
order in this theory. The usual renormalization counterterms 
cancel the ultraviolet divergences associated with mass, coupling 
and wave-function, and the Bogoliubov coefficients are chosen 
consistently to cancel the initial time singularities. That is, their
high energy behaviour is fixed according to the initial data
[sec. IV]. 

We have established contact with linear response theory  
by obtaining the evolution equations for the scalar condensate 
and the initial value problem in the linearized approximation as 
the linear response to an external source coupled to the scalar 
condensate. This equivalent formulation clarifies at once the 
relationship between the linearized approximation for the evolution 
equations of the condensate and linear response. The corresponding
initial value problem, i.e. providing Cauchy data for the field and 
its first derivative on a spatial hypersurface  requires that 
the external source that couples to the scalar field {\em  does not vanish 
after the initial time}. A very specific source term that depends on a
given past history of the condensate furnished a singularity free
initial value problem.  

After presenting the method in the simpler homogeneous case and establishing 
the relationship to linear response theory
we focused on the important case of {\bf inhomogeneous} condensates. 
Following on the steps for the homogeneous case we
have constructed the proper Bogoliubov transformation to lowest order 
in the Yukawa coupling and shown explicitly how a judicious choice of the
Bogoliubov coefficients in combination with the usual 
renormalization counterterms leads to an initial value problem free
of ultraviolet and initial time singularities. As an example of this 
consistent procedure we have provided a numerical study of the 
space-time evolution of an inhomogeneous scalar condensate both 
in the case in which the scalar can decay into fermion-antifermion 
pairs and in the case in which the scalar is light and stable. 

A noteworthy result of our study is the breakdown of (discrete) 
chiral symmetry via the {\em initial conditions} in the case of 
massless fermions. The initial state with a non-zero value of the 
condensate of its time derivatives (first or second) break the 
discrete chiral symmetry, we emphasize that this breakdown is 
not a consequence of an explicitly symmetry breaking term but of 
the initial quantum state with a non-equilibrium condensate. 
This is clearly a manifestation of non-equilibrium effects.

We have also generalized the results of the fermionic case to 
scalar field theories by exploiting the relation to linear response, 
thus providing a generalized and consistent manner of describing 
non-equilibrium evolution of condensates in terms of an initial 
value problem free of ultraviolet divergences and initial time singularities.

{\bf Applications:} We foresee several applications of these 
methods: i) we can now study consistently the evolution of 
inhomogeneous pion condensates after a chiral phase transition by 
setting up a physically reasonable initial value problem that 
incorporates the important features of the transition in 
the {\em fully interacting} initial state (or density matrix). 
This approach is complementary to that advocated in 
ref.\cite{boycoopveg}. 
ii) In cosmology we can now study the non-equilibrium dynamics 
of inhomogeneous configurations by
providing the initial field profile and the first derivative 
on a space-like hypersurface and following the space-time
evolution of this configuration. In particular a very relevant 
setting for cosmology is that of supersymmetric theories during 
for example the stages of rolling of the scalar field component. Treating
the dynamics as an initial value problem, the initial conditions 
on the scalar field, displaced from the equilibrium position breaks 
supersymmetry. This breakdown of supersymmetry is {\em not} explicit 
at the level of the Lagrangian, but by the quantum state. Our 
formulation allows us to follow the dynamics consistently and 
study the consequences of this supersymmetry breaking. Work on 
these issues is in progress. 

The next step in our program is to extend the results obtained in 
this article, valid in the linearized approximation, to
a full non-linear inhomogeneous problem. We expect to report on 
progress on these and other issues in the near future.

\section{Acknowledgements} 
J. B. thanks the theory groups of the University of
Pittsburgh and of the LPTHE at the Universit\'e Pierre et Marie Curie,
Paris for their hospitality, and the Deutsche Forschungsgemeinschaft 
for partial support under grant Ba-703/6-1. D. B. thanks PHY-9605186,
INT-9815064 for partial support, and the LPTHE at the Universit\'es
Pierre et Marie Curie and Denis Diderot (Paris VI and VII) for their 
hospitality.  D.B. and H. J. de V. acknowledge support from NATO.  

\newpage
\begin{appendix}
\section{Fermionic Green's functions in the Bogoliubov state for the
homogeneous case} 
\setcounter{equation}{0}

The Green function $ S^>_{\flat}(t,\bfx;t',\bfx') $ in the
Bogoliubov-transformed state $ |0_{\flat} \rangle $ is defined via
\be\label{funGtra}
i S^>_{\flat}(t,\bfx;t',\bfx')=\langle
 0_{\flat}|\psi(t,\bfx)\bar\psi(t',\bfx') | 0_{\flat} \rangle
\ee
and the (free) field operators are given, in terms of the
creation and annihilation operators,  by
\bea
\psi(t,\bfx)&=&\sum_s\intp \;  e^{i{\bf p\bf x}}
\left[ b(\bfp,s) \; U(\bfp,s) \; e^{-i\Ep t}\right. \nonumber \\
&&\left.\hspace*{30mm}+d^\dagger(-\bfp,s) \; 
V(-\bfp,s) \; e^{i\Ep t}\right] \nonumber \\
\bar\psi(t',\bfx')&=&\sum_{s'}\intpp  \; e^{-i\bfp'\bfx'}
\left[b^\dagger(\bfp',s') \; \bar U(\bfp',s') \; e^{-i\Epp t'}\right.
 \nonumber \\ && \hspace*{30mm}\left.
+ \; d(-\bfp',s')  \; \bar V(-\bfp',s') \; e^{i\Epp t'}\right]
\pkt \label{psipsiB}
\eea
We use the normalization for the spinors $ \bar U(\bfp,s) U(\bfp,s)
=2m $ so that $ U^\dagger(\bfp,s)U(\bfp,s)=1 $.

Using eqs.(\ref{bdbemol}), (\ref{mdens}), (\ref{funGtra}) and
(\ref{psipsiB}) we obtain the following expression for  the
transformed Green function $S^>_{\flat}(t,\bfx;t',\bfx') $,
\bea &&i
S^>_{\flat}(t,\bfx;t',\bfx')=\langle0_{\flat}|\psi(t,\bfx)\bar\psi(t',\bfx')
|0_{\flat}\rangle= \nonumber\\ \nonumber
&&\sum_s\intp e^{i\bfp(\bfx-\bfx')}
\left[\cos^2(\beta_{ps}) \; U(\bfp,s)\bar U(\bfp,s) \;e^{-i\Ep(t-t')} 
\right..
\\ \nonumber &&
-\sin\beta_{ps} \;\cos\beta_{ps} \;e^{i\delta_{ps}} \;U(\bfp,s)\bar
V(-\bfp,s)  \;
e^{-i\Ep(t+t')}\\ \nonumber &&
-\sin\beta_{ps} \;\cos\beta_{ps} \;e^{-i\delta_{ps}}V(-\bfp,s)\bar
U(\bfp,s)  \;
e^{i\Ep(t+t')}\\ \nonumber &&
\left. +\sin^2\beta_{ps} \;V(-\bfp,s) \;\bar V(-\bfp,s) \;e^{i\Ep(t-t')}
\right]\pkt\eea

As long as we consider homogeneous condensates, the 
angles $\beta_{ps}$ and $\delta_{ps}$ can be chosen to depend  only on the
modulus  $ |\bfp| $, and on the helicities. The
weight of the two possible helicities is still arbitrary and we consider
these angles to be functions of the helicity matrix
\bea\nonumber
\bSigma \hat \bfp =
\left(\begin{array}{cc}
\bsigma \hat \bfp & 0 \\ 0
& \bsigma \hat \bfp \end{array}\right )
\kma\eea
where $ \hat \bfp = \bfp/|\bfp| $.
We have, e.g., 
\bea\nonumber
f(\bSigma\hat \bfp)\sum_s U(\bfp,s) \bar V(-\bfp,s) =
\sum_s U(\bfp,s) \bar V(-\bfp,s)f(\bSigma\hat\bfp)
=\sum_s f(s) U(\bfp,s) \bar V(-\bfp,s)
\pkt \eea
$\bSigma\hat\bfp$ commutes with the all other matrices
that are relevant to this discussion, 
$\gamma_0,\gamma_5$ and $\bfp\bgamma$ and  can therefore be treated 
as a c-number.

Specifying the vector $p^\mu$ to be on shell, $p^\mu = (\Ep, \bfp)$, and
using $\gamma_0 \bgamma \gamma_0 = -{\bf \bgamma}$
it is straightforward to find
\bea\nonumber
\sum_s U(\bfp,s)\bar U(\bfp,s) &=& \slash\hspace{-2.3mm} p +m \kma \\ 
\sum_s V(-\bfp,s)\bar V(-\bfp,s) &=& \gamma_0 (\slash
\hspace{-2.3mm} p -m)\gamma_0 = \gamma_5\gamma_0(\slash\hspace{-2.3mm}
p+m)\gamma_5\gamma_0\pkt \nonumber\eea
\noindent the mixed product can be found by resorting to the representation
\bea\nonumber
U(\bfp,s) = \frac{\slash\hspace{-2.3mm}p +m}
{\sqrt{\Ep+m}}\left(\begin{array}{c}\chi_s \\ 0 \end{array}
\right)\\
V(-\bfp,s) = - \frac{\gamma_0(\slash\hspace{-2.3mm}p -m)\gamma_0}
{\sqrt{\Ep+m}}\left(\begin{array}{c}0\\ \chi_s
\end{array}\right)\pkt\nonumber
\eea
Where $ \chi_s $ is an eigenspinor of $\bsigma \hat \bfp $ with
eigenvalue $ s $ from which we find
\bea\nonumber
\sum_s V(-\bfp,s)\bar U(\bfp,s)&=&\gamma_5 \gamma_0 (\slash \hspace{-2.3mm}p
+m) \kma\\
\sum_s U(\bfp,s)\bar V(-\bfp,s)&=&(\slash\hspace{-2.3mm}p+m)\gamma_5\gamma_0
\kma \nonumber\eea

We note that the phase $\delta_{ps}$ appears in a combination such that
a shift in this phase can be compensated by a shift in the origin of time,
i.e. a time translation $t\rightarrow t+t_o$ $\delta_{ps} \to \delta_{ps} +
2 \Ep t_0$. Since the square of the helicity matrix is the identity and
the only odd function of $\beta_{ps}$ multiplies the mixed terms, we found
that the simplest Bogoliubov transformation that is required to cancel
the initial time singularities is such that the phase $ \delta_{ps} $ is
independent of $ s $ and that $ \beta_{ps} = \bSigma\hat \bfp  \beta_p $
so that  
\bea\nonumber
\cos \beta_{ps} = \cos \beta_p \\
\sin \beta_{ps} = \bSigma\hat \bfp \sin \beta_p
\pkt \nonumber\eea

Such a choice has proven to be appropriate for removing the
initial singularity for the full one-loop equations.

After some straightforward algebra, we find 
\bea \label{sbmod}
&&i S^>_{\flat}(t,\bfx;t',\bfx')=\langle 0_{\flat}
|\psi(t,\bfx)\bar\psi(t',\bfx') | 0_{\flat}\rangle= 
\\ \nonumber
&&\intp e^{i\bfp(\bfx-\bfx')}
\left[\cos\beta_p \; e^{-i\Ep t}-
\sin\beta_p  \; e^{-i\delta_p} \;\bSigma\hat\bfp
\gamma_5\gamma_0 \; e^{i\Ep t} \;\right]
\\ \nonumber && 
(\slash\hspace{-2.3mm}p+m)\left[\cos\beta_p  \; e^{i\Ep t'}-
\sin\beta_p \;e^{i\delta_p}
\gamma_5\gamma_0 \;\bSigma\hat\bfp \; e^{-i\Ep t'}\right]
\kma
\eea

Upon reordering of the terms we find the Green's function quoted in
(\ref{Sgreatbogo}). 

A similar calculation leads to the transformed Green function 
$ S^<_{\flat}(t,\bfx;t',\bfx') $ which is defined as
\bea \nonumber
-i S^<_{\flat}(t,\bfx;t',\bfx')=
\langle 0_{\flat}|\bar\psi(t',\bfx')\psi(t,\bfx)
| 0_{\flat} \rangle
\kma\eea

Following the same steps leading to (\ref{sbmod}) we find
\bea  \nonumber
&&-i S^<_{\flat}(t,\bfx;t',\bfx')=\langle 0_{\flat}
|\bar\psi(t',\bfx')\psi(t,\bfx) | 0_{\flat}\rangle= 
\\ \nonumber
&&\intp e^{i\bfp(\bfx-\bfx')}
\left[\sin\beta_p \; e^{i\delta_{ps}}\;e^{-i\Ep t}\;\bSigma\hat\bfp+
\cos\beta_p\;\gamma_5\gamma_0\; e^{i\Ep t}\right]
\\ \nonumber && 
(\slash\hspace{-2.3mm}p+m)
\left[\sin\beta_p \;e^{-i\delta_p}\;e^{i\Ep t'}\;\bSigma\hat\bfp+
\cos\beta_p\;\gamma_5\gamma_0\; e^{-i\Ep t'}\right]
\pkt
\eea

which upon reordering of terms gives the form quoted in 
expression (\ref{Ssmallbogo}).

\section{Bogoliubov transformation and tadpole diagram for
inhomogenous systems}
\setcounter{equation}{0}
In this Appendix we generalize the Bogoliubov transformations described
in the homogeneous case to the case of inhomogeneous condensates.
Unlike the homogeneous case in which the generator of the Bogoliubov 
transformation creates particle-antiparticle pairs of zero total momentum,
in the inhomogenous case the total momentum of the pair is non-zero. 

Consistent with perturbation theory we now find the corresponding 
Bogoliubov transformation to lowest order in the Yukawa coupling, 
thus $\cos \beta_{ps} \approx 1 ~, ~ \sin \beta_{ps} \approx 
\beta_{ps} = {\cal O}(g) $. 
Since the self-energy is already of ${\cal O}(g^2)$, to lowest order we 
only need to focus on the tadpole term $J_{\flat}({\bf x},t)$. 

The Bogoliubov transformation in lowest order reads
\bea \nonumber
 b(\bfp,s)&=&
  b_{\flat}(\bfp,s)+\intpp \rho_{ss'}(\bfp,\bfp')\;
  d^\dagger_{\flat}(-\bfp',s')
\kma \\ \nonumber
 d^\dagger(-\bfp,s)&=&
-\intpp \rho_{s's}^*(\bfp',\bfp) b_{\flat}(\bfp',s')
+  d^\dagger_{\flat}(-\bfp,s)
\pkt
\eea
\noindent with $ \rho_{ss'} = {\cal O}(g) $. 
With this choice  the  transformation leaves the
 canonical anticommutation relations unchanged up to terms of order
$ \rho_{ss'}^2 $. In order to compute the transformed
Green functions we need  the expectation values
of bilinear combinations of creation and annihilation operator.
We find the following expectation values that are necessary to compute the
Green's functions , 
\bea \nonumber
&&\langle  0_{\flat}|b(\bfp,s)d(-\bfp',s')| 0_{\flat}\rangle
\\ \nonumber
&&=\intppp\left[-\rho_{s''s'}(\bfp'',\bfp')\right]
\langle  0_{\flat}| b_{\flat}(\bfp,s)
 b_{\flat}^\dagger(\bfp'',s'')| 0_{\flat}\rangle
\delta_{ss'}(2\pi)^3 2\Ep\delta^3(\bfp-\bfp'')
\\ 
&&=-\rho_{ss'}(\bfp,\bfp')
\pkt  \nonumber\eea

\bea \nonumber
\langle  0_{\flat} |b(\bfp,s)b^\dagger(\bfp',s')|0_{\flat}\rangle
&=&(2\pi)^3 2 E_{\bf p} \delta_{ss'}\delta^3(\bfp-\bfp')
\kma \\ \nonumber
\langle 0_{\flat}|d^\dagger(-\bfp,s)d(-\bfp',s')|0_{\flat}\rangle
&=&0
\kma \\ \nonumber
\langle 0_{\flat}|d^\dagger(-\bfp,s)b^\dagger(\bfp',s')|0_{\flat}\rangle
&=& -\rho^*_{s's}(\bfp',\bfp)\pkt \\
\langle 0_{\flat}|d(\bfp',s')d^\dagger(\bfp,s)|0_{\flat}\rangle
&=&(2\pi)^3 2 E_0\delta_{ss'}\delta^3(\bfp-\bfp')
\kma \nonumber \\ \nonumber
\langle 0_{\flat}|b^\dagger(-\bfp',s')b(-\bfp,s)|0_{\flat}\rangle
&=&0
\kma \\ \nonumber
\langle 0_{\flat}|d(-\bfp',s')b(\bfp,s)|0_{\flat}\rangle
&=& \rho_{ss'}(\bfp,\bfp')
\kma \\ \nonumber
\langle 0_{\flat}|b^\dagger(\bfp',s')d^\dagger(-\bfp,s)|0_{\flat}\rangle
&=&\rho^*_{s's}(\bfp',\bfp)
\pkt
\eea
This yields   the transformed Green's functions,
\bea \nonumber
&&i
S_{\flat}^>(t,\bfx;t',\bfx')=\langle 0_{\flat}|\psi(t,\bfx)\bar\psi(t',\bfx') 
|0_{\flat}\rangle= 
\\ \nonumber
&&\sum_s\intp e^{i\bfp(\bfx-\bfx')} \;
U(\bfp,s) \; \bar U(\bfp,s) \; e^{-i\Ep(t-t')} 
\\ \nonumber &&-\sum_{ss'} \intp \intpp  \;
e^{i(\bfp\bfx-\bfp'\bfx')} \; \left\{\rho^*_{s's}(\bfp',\bfp)  \;
V(-\bfp,s) \; \bar U(\bfp',s') \; e^{-i(\Ep t+\Epp t')}\right.\\ \nonumber
&&\hspace{20mm}
\left.+\rho_{ss'}(\bfp,\bfp') \; 
U(\bfp,s) \; \bar  V(-\bfp',s') \; e^{i(\Ep t+\Epp t')}\right\}
\pkt \eea 
       
and 

\bea \nonumber
&&-i S_{\flat}^<(t,\bfx;t',\bfx')=
\langle0_{\flat}|\bar\psi(t',\bfx')\psi(t,\bfx)|0_{\flat}\rangle= 
\\ \nonumber
&&\sum_s\intp e^{i\bfp(\bfx-\bfx')} \;
V(-\bfp,s) \; \bar V(-\bfp,s) \; e^{-i\Ep(t-t')}  \;
\\ \nonumber &&+\sum_{ss'} \intp \intpp  \;
e^{i(\bfp\bfx-\bfp'\bfx')}\left\{\rho^*_{s's}(\bfp',\bfp)  \;
V(-\bfp,s) \;\bar U(\bfp',s') \;e^{-i(\Ep t+\Epp t')}\right.\\ \nonumber
&& \left.+ \;\rho_{ss'}(\bfp,\bfp')  \;
U(\bfp,s) \;\bar V(-\bfp',s') \;e^{i(\Ep t+\Epp t')}\right\}
\pkt \eea 
to first order in $ \rho_{ss'}(\bfp,\bfp') $.

Using these results we now can evaluate the tadpole graph
in the inhomogenous condensate. That is, the expectation value of 
 $ < \bar\psi \psi> $ which plays the r\^ole of  an external current in
 the equation of motion. Inserting the above explicit expressions we find
\bea  \nonumber
&&\hspace{-20mm}J_\flat(t,\bfx)= ig\tr S_{\flat}^>(t,\bfx;t,\bfx)=
g \sum_s\intp \tr \left[U(\bfp,s)\bar U(\bfp,s)\right] 
\\ \nonumber &&-\sum_{ss'} \intp \intpp 
e^{i(\bfp-\bfp')\bfx}\\ \nonumber
&&\left\{\rho^*_{s's}(\bfp',\bfp) 
\tr \! \left[V(-\bfp,s) \; \bar U(\bfp',s')\right]
e^{i(\Ep +\Epp)t }\right.\\ \nonumber
&&
\left.\rho_{ss'}(\bfp,\bfp') 
\tr \! \left[U(\bfp,s) \; \bar V(-\bfp',s')\right]
e^{-i(\Ep +\Epp) t}\right\}\label{jota}
\pkt \eea
The traces over the  spinors yield,
\bea
 \tr V(-\bfp,s)\bar U(\bfp',s')&=&\bar U(\bfp',s') V(-\bfp,s)
\\  \nonumber
&=&-\delta_{ss'}s\left[\sqrt{(\Ep-m)(\Epp+m)}+\sqrt{(\Ep+m)(\Epp-m)}\right]
\\  \nonumber
 \tr U(\bfp,s)\bar V(-\bfp',s')&=&\bar V(-\bfp',s') U(\bfp,s)
\\ \nonumber 
&=&-\delta_{ss'}s\left[\sqrt{(\Ep-m)(\Epp+m)}+\sqrt{(\Ep+m)(\Epp-m)}\right]
\pkt\eea
We see that the relevant part of $ \rho_{ss'}(\bfp,\bfp') $
contributing to eq.(\ref{jota}) is odd in $s$ and diagonal in $ ss'
$. We  therefore choose,
\bea \nonumber
\rho_{ss'}(\bfp,\bfp')= \gamma(\bfp,\bfp')
\frac{s\; \delta_{ss'}}{\sqrt{(\Ep-m)(\Epp+m)}+\sqrt{(\Ep+m)(\Epp-m)}}
\pkt\eea
Then,
\bea \nonumber
&&J_\flat(t,\bfx)= 4 mg\intp  
\\ \nonumber &&-2 \intp \intpp e^{i(\bfp-\bfp')\bfx}\left\{
\gamma^*(\bfp',\bfp)  \; e^{-i(\Ep +\Epp)t} +\gamma(\bfp,\bfp')  \;
e^{i(\Ep +\Epp) t}\right\}
\pkt \eea 

The first term is again space and time independent
and is absorbed into a shift of the condensate. The second term displays 
space and time dependence in a factored form. It will be used 
to compensate for the initial time singularities of the self-energy.

\section{Analysis of the self energy kernel}
\setcounter{equation}{0}
In this Appendix we provide the details for the various contributions to
the self-energy. The  integrals that enter in the expression for 
the self energy kernel can be related to the following one 
defined in dimensional regularization
\bea \nonumber
&&I(q_0^2,\bfq^2)\equiv
i\int\frac{d^{3-\epsilon}p}{(2\pi)^{3-\epsilon}}
\frac{1}{2E_+E_-}\frac{E_+ +E_-}{\left(E_+ +E_-\right)^2-q_0^2}
\\ \nonumber
&&=\int \frac{d^{4-\epsilon}p}{(2\pi)^{4-\epsilon}}
\frac{1}{\left[\left(p-q/2\right)^2-m^2+io\right]
\left[\left(p+q/2\right)^2-m^2+io\right]}
\\ \nonumber
&&=\frac{1}{16\pi^2}\left[L_\epsilon
+\int_0^1 d\alpha \ln\frac{m^2}{m^2+\alpha(1-\alpha)\left(\bfq^2
-q_0^2\right)}\right]\pkt
\eea
Where we have introduced the shifted momenta
$\bfp_\pm=\bfp\pm \bfq/2$ and  energies $E_\pm=\sqrt{\bfp_\pm^2+m^2}$.
$L_\epsilon$ is defined as
\bea \nonumber
L_\epsilon=\frac{2}{\epsilon}-\gamma+\ln\frac{4\pi\mu^2}{m^2}
\pkt\eea
We now consider the various integrals defined in section
VIII. In doing so we will shift the integration variable
$\bfp$ so that $\bfp\to\bfp_+=\bfp+\bfq/2$ and
$\bfp'=\bfp-\bfq \to \bfp_-=\bfp-\bfq/2$. Then
the numerator arising from the Dirac trace takes the form
\bea \nonumber
E_+E_-+\bfp_+\bfp_--m^2=\frac{1}{2}(E_++E_-)^2
-2\left(m^2+\frac{\bfq^2}{4}\right)
\eea
We then have
\bea \nonumber
\Sigmat_1(\bfq^2)=-8g^2 I_1(\bfq^2)
\eea
with
\bea\nonumber
I_1(\bfq^2)&=&\int\frac{d^{3-\epsilon}p}
{(2\pi)^{3-\epsilon}}\frac{E_+E_-+\bfp_+\bfp_--m^2}
{4E_+E_-\left(E_++E_-\right)}
\\ \nonumber
&=&\frac{1}{8}\int\frac{d^{3-\epsilon}p}
{(2\pi)^{3-\epsilon}}\left(\frac{1}{E_+}+\frac{1}{E_-}\right)
-\left(m^2+\frac{\bfq^2}{4}\right)\int\frac{d^{3-\epsilon}p}
{(2\pi)^{3-\epsilon}}\frac{1}{2E_+E_-(E_++E_-)}
\eea
The first integral, including the prefactor, is equal to
\bea\nonumber
\int\frac{d^{3-\epsilon}p}
{(2\pi)^{3-\epsilon}}\frac{1}{4E}=-\frac{m^2}{32\pi^2}(L_\epsilon+1)
\pkt\eea
The second integral is  the basic integral
$I(q_0^2,\bfq^2)$ at $q_0^2=0$.
Altogether we obtain
\bea\nonumber
\Sigmat_1(\bfq^2)=\Sigmat_1(0)+\bfq^2\Sigmat'_1(0)+\Delta\Sigmat_1(\bfq^2)
\eea
with
\bea\nonumber
\Sigmat_1(0)&=&-\delta M^2
=\frac{3m^2g^2}{4\pi^2}(L_\epsilon+\frac{1}{3})\\
\Sigmat'_1(0)&=&-\delta Z = \frac{g^2}{8\pi^2}L_\epsilon \label{delZ}
\\ \Delta\Sigmat_1(\bfq^2)&=&\left(m^2+\frac{\bfq^2}{4}\right)
\frac{g^2}{2\pi^2}
\int_0^1 d\alpha \ln\frac{m^2}{m^2+\alpha(1-\alpha)\,\bfq^2} 
\pkt\nonumber
\eea
Here we have introduced the renormalization constants corresponding
to a renormalization at $q^2=0$.

For $ \Sigmat_3(\bfq^2) $ we have
\bea\nonumber
\Sigmat_3(\bfq^2)=8 \; g^2 \; I_3(\bfq^2)
\eea
with
\bea\nonumber
I_3(\bfq^2)=\int\frac{d^{3-\epsilon}p}
{(2\pi)^{3-\epsilon}}\frac{1}{4E_+E_-}
\frac{(E_++E_-)^2/2-2(m^2+\bfq^2/4)}{(E_++E_-)^3}
\pkt\eea
This integral can be related to the integral $I(q_0^2,\bfq^2)$ and its
derivative w.r.t. $q_0^2$, at $q_0=0$. We find 
\bea\nonumber
\Sigmat_3(\bfq^2)=-\delta Z +\Delta\Sigmat_3(\bfq^2)
\eea
where $\delta Z$ has been defined in eq.\ref{delZ}). The finite part is
\bea\nonumber
\Delta \Sigmat_3(\bfq^2)
&=&-\frac{g^2}{8\pi^2}\int_0^1d\alpha \ln\left[1+\alpha(1-\alpha)
\frac{\bfq^2}{m^2}\right] \\ \nonumber
&&+\frac{g^2}{2\pi^2}
\left(m^2+\frac{\bfq^2}{4}\right)\int_0^1  d\alpha
\frac{\alpha(1-\alpha)}{m^2+\alpha(1-\alpha)\bfq^2}
\eea
From the way in which we have introduced
$\delta Z$ in $\tilde\Sigma_1$ and $\tilde\Sigma_3$ it is apparent
that the covariant counterterms 
$\delta Z (\ddot \phit+\bfq^2\phit)$
in the equation of motion will absorb these divergences.

We finally consider the Laplace transform of
the subtracted self energy kernel introduced in eq.(\ref{self_lap}) 
\bea\nonumber
\sigmat_s(s^2,\bfq^2)=-8 \;g^2 \; s^2\int\frac{d^3p}{(2\pi)^3}
\frac{1}{4E_+E_-} \frac{(E_++E_-)^2/2-2(m^2+\bfq^2/4)}{(E_++E_-)^3
[(E_++E_-)^2+s^2]}
\pkt\eea
Comparing with the standard integral $I(q_0^2,\bfq^2)$
we see that besides the continuation to the Euclidean
region, $q_0^2\to -s^2$ we have additional denominators.
These can be obtained via subtraction. We have
\bea\nonumber
&&\int\frac{d^3p}{(2\pi)^3} \frac{1}{2E_+E_-}
\frac{1}{(E_++E_-) [(E_++E_-)^2+s^2]}=-\frac{1}{s^2}\left[I(-s^2,\bfq^2)
-I(0,\bfq^2)\right] \\ 
&&=-\frac{1}{s^2}\frac{1}{16\pi^2}
\int_0^1 d\alpha \; \ln\frac{m^2+\alpha(1-\alpha)\bfq^2}
{m^2+\alpha(1-\alpha)(\bfq^2+s^2)}
\pkt\nonumber\eea
We proceed analogously for the second integral
and finally obtain
\bea\nonumber
\sigmat_s(s^2,\bfq^2)&=&\frac{g^2}{2\pi^2} \int_0^1 d\alpha
\left\{\frac{1}{s^2} \left[m^2+\frac{1}{4}(\bfq^2+s^2)\right]
\ln\frac{m^2+\alpha(1-\alpha)\bfq^2}
{m^2+\alpha(1-\alpha)(\bfq^2+s^2)} \right. \\ \label{self_lap_ex}
&&\left.+\left(m^2+\frac{\bfq^2}{4}\right)
\frac{\alpha(1-\alpha)}{m^2+\alpha(1-\alpha)\bfq^2}
\right\}
\eea
The $\alpha$ integrations can be performed analytically.

\section{Details of the analytic solution}
\setcounter{equation}{0}
We have obtained in section  the solution of the equation of motion 
and its solution via Laplace transform.
We consider at first the unrenormalized equation.
The solution reads
\bea\nonumber
 \psi_{\bf q} (s) =\frac{ s \; {\phi}_{\bf q}(0)+\dot
{\phi}_{\bf q}(0)}{s^2+M^2+\bfq^2+\tilde \sigma(s^2,\bfq^2)}
\kma\eea
so that 
\bea\nonumber
\phi_{\bf q}(t)=\frac{1}{2\pi i}\int_{-i\infty+c}^{i\infty +c}
ds\, e^{st}\; 
\frac{s {\phi}_{\bf q}(0)+\dot{\phi}_{\bf q}(0)}
{s^2+M^2+\bfq^2+\tilde \sigma(s^2,\bfq^2)}
\pkt\eea
As usual \cite{boyinho,boyareheat} we shift the contour to the left
so that finally it includes the cuts, and eventually poles, on the
imaginary $s$ axis and a circle at $|s|\to \infty$ around 
the left half, which does not contribute for positive $t$
 as the exponential
$\exp st$ tends to zero there.
In doing so we make use of the causality condition that 
there are no zeros in the left half of the complex $s$ plane, as
required by causality.
Along the cut at $s=i\omega$ with 
$2\sqrt{m^2+\bfq^2}<\omega<\infty$
 we define the real and imaginary parts
of the kernel $\sigmat$ by the convention
\bea\nonumber
\sigmat((i\omega\pm \epsilon)^2,\bfq^2)=
\sigmat_R(-\omega^2,\bfq^2)\pm i \;  \sigmat_I(-\omega^2,\bfq^2)
\pkt\eea
As $\sigma$ only depends on $s^2$ this also fixes the 
relative signs
of the imaginary parts of $\sigmat$  on the lower cut for which
$-\infty < \omega < -2 \sqrt{m^2+\bfq^2}$. 
We then obtain
\bea\nonumber
&&{\phi}_{\bf q}(t)= \frac{1}{2\pi i} \int_{\omega_c}^{\infty}
i \; d\omega  \; e^{i\omega t} \;
\disc\,\frac{i\omega\phi_{\bf q}(0)+\dot\phi_{\bf q}(0)}
{-\omega^2+M^2+\bfp^2+\sigmat(-\omega^2\pm i\epsilon,\bfq^2)}
\nonumber \\
&&+\frac{1}{2\pi i}\int_{\omega_c}^{\infty}
(-i) \;d\omega\;  e^{-i\omega t} \;
\mbox \disc\,\frac{-i\omega\phi_{\bf q}(0)+\dot{\phi}_{\bf q}(0)}
{-\omega^2+M^2+\bfp^2+\sigmat(-\omega^2\mp i\epsilon,\bfq^2)}\nonumber
\eea
with $\omega_c=2\sqrt{\bfq^2+m^2}$ for the two fermion cut.
The spectral density is obtained from the discontinuity across the cut 
\begin{eqnarray}\nonumber
S(\omega,{\bf q}) & = & i \; \disc \,\frac{1}
{-\omega^2+M^2+\bfp^2+\sigmat(-\omega^2 + i\epsilon,\bfq^2)} \nonumber \\
& = &
\frac{2\sigmat_I(-\omega^2,\bfq^2)}
{\left[-\omega^2+M^2+\bfp^2+
\sigmat_R(-\omega^2+ i\epsilon,\bfq^2)\right]^2+\sigmat^2_I(-\omega^2+ i\epsilon,\bfq^2)}
\end{eqnarray}

In the case in which the scalar particle is unstable, i.e. $M>2m$
there is a resonance above the fermion-antifermion 
threshold and no support for the spectral density below
threshold\cite{boyinho,boyareheat}. However in the case $M<2m$ the
scalar is stable and cannot decay, now the spectral density has
support above and {\em below} threshold. Below threshold 
the spectral density is a delta function at the position of the
renormalized pole, to include the stable pole below 
the two particle threshold in the description we now define
$\omega_c=0^+$ to distinguish that the origin is excluded 
from the integration region. The pole in the stable case  is obtained
from the identity 
\bea\nonumber
S(\omega,{\bf q}) \stackrel{\sigma_I \rightarrow 0}{\rightarrow} \pi
\delta(-\omega^2+M^2+\bfp^2+ 
\sigmat_R(-\omega^2+ i\epsilon,\bfq^2)) 
\eea
so that
\bea\nonumber
&&\phi_{\bf q}(t)=
\frac{2}{\pi}\int_{0^+}^{\infty} d\omega  \;\cos\omega t \;
\frac{\omega \; \phi_{\bf q}(0)\; \sigmat_I(-\omega^2,\bfq^2)}
{\left[-\omega^2+M^2+\bfp^2+
\sigmat_R(-\omega^2+ i\epsilon,\bfq^2)\right]^2+\sigmat^2_I(-\omega^2+
i\epsilon,\bfq^2)} 
 \nonumber\\
&&+\frac{2}{\pi}\int_{0+}^{\infty}
d\omega \; \sin\omega t \;
\frac{\dot\phi_{\bf q}(0)\; \sigmat_I(-\omega^2,\bfq^2)}
{\left[-\omega^2+M^2+\bfp^2+\sigmat_R(-\omega^2+ i\epsilon,\bfq^2)\right]^2+\sigmat^2_I(-\omega^2+ i\epsilon,\bfq^2)}
\pkt\eea
In order that this equation and its time derivative
be consistent at $ t=0 $ we have to require the sum rule
\bea\nonumber
1=\frac{2}{\pi}\int_{0^+}^{\infty} d\omega  \; \frac{\omega\;
\sigmat_I(-\omega^2,\bfq^2)} {\left[-\omega^2+M^2+\bfp^2+
\sigmat_R(-\omega^2+ i\epsilon,\bfq^2)\right]^2+\sigmat^2_I(-\omega^2+
i\epsilon,\bfq^2)}  
\pkt\eea
In order to derive this sume rule we require, as
already mentioned above, that the denominator 
$s^2+\bfq^2+M^2+\sigma(s^2,\bfq^2)$ has no zeros
in the left half of the complex plane. We have to assume
furthermore that $\tilde\sigma(s^2,\bfq^2)$ increases less
strongly as $s^2$ as $|s|\to \infty$ in the left half of the
complex plane.  Under these assumptions we have 
the identity
\bea\nonumber
\frac{1}{i\pi}\oint ds \;
\frac{s}{s^2+\bfq^2+M^2+\sigmat(s^2,\bfq^2)}=0
\eea
if the integral is carried out along the contour enclosing
the left half of the complex plane. The contour consists of an
integral along the left of the imaginary $s$ axis and a semicircle at
$|s|=\infty$. The latter one contibutes
\bea\nonumber
\frac{1}{i\pi}\int_{\sf C}\frac{ds}{s}=-1
\pkt\eea
The integral along the imaginary axis is given by
\bea\nonumber
&&\frac{1}{\pi}\int_0^\infty d\omega \;
\left[\frac{i\omega}{s^2+\bfq^2+M^2+\sigmat(-\omega^2
-io)}+\frac{-i\omega}{s^2+\bfq^2+M^2+\sigmat(-\omega^2
+io)}\right]
\nonumber \\
&&=\frac{2}{\pi}\int_{\omega_c}^{\infty} d\omega  \;
\frac{\omega \tilde{\sigma}_I(-\omega^2,\bfq^2)}
{\left|-\omega^2+M^2+\bfp^2+
\tilde \sigma(-\omega^2+ i\epsilon,\bfq^2)\right|^2} 
\pkt\eea
The two parts of the contour integral have to add up to zero, which
yields the sum rule. 

For the renormalized equation of motion we rewrite the result obtained
in section \ref{solu} as 
\bea\nonumber
\phi_{\bf q}(t)=\frac{1}{2\pi i}\int_{-i\infty+c}^{i\infty+c}ds  \;
\left\{\left[s\phi_{\bf q}(0)+\dot{\phi}_{\bf q}(0)\right]
F_1(s^2,\bfq^2)  +\ddot\phi_{\bf q}(0)\frac{1}{s}
F_2(s^2,\bfq^2)\right\}
\eea
with
\bea\nonumber
F_1(s^2,\bfq^2)&=&\frac{1+\Delta\Sigmat_3(\bfq^2)+s\; 
\tilde\sigma(s^2,\bfq^2)}{s^2\left[1+\Delta\Sigmat_3(\bfq^2)+
\sigmat_s(s^2,\bfq^2)\right]+\bfq^2+M^2+\Delta\Sigmat_1
(s^2,\bfq^2)}\\\nonumber
F_2(s^2,\bfq^2)&=&\frac{\sigmat_s(s^2,\bfq^2)}
{s^2\left[1+\Delta\Sigmat_3(\bfq^2)+
\sigmat_s(s^2,\bfq^2)\right]+\bfq^2+M^2+\Delta\Sigmat_1
(\bfq^2)}  
\pkt\eea
The function $F_1 (s^2,\bfq^2) $ has analyticity properties
analogous to the fraction $ 1/(s^2+\bfq^2+M^2+\tilde \sigma) $
considered above. In particular the discontinuities along
the positive and negative imaginary axis have the same
relative signs, it has no singularities in the left half 
of the complex plane, and the limiting behavior
as $ |s|\to \infty $ is $ 1/s^2 $. For the first property it is essential
to note that $\sigmat_s$ only depends on the square of the
variable $ s $, see (\ref{self_lap_ex}).
For the last property is sufficient to notice that
 $ \tilde\sigma_s(s^2,\bfq^2) $ behaves as $ \ln s^2 $ as 
$ |s|\to \infty $, so the terms proportional to
 $ \tilde \sigma_s $ dominate in numerator and denominator.
The function $ F_2(s^2,\bfq^2) $ has analyticity properties
analogous to those of $ F_1(s^2,\bfq^2) $, and decreases asymptotically
as $1/s^2$.  The relative signs of the
imaginary parts along the cuts are the same as for
$ F_1(s^2,\bfq^2) $. Collecting all terms we find 
\bea\label{finally}
\phi_{\bf q}(t)&=&\frac{2}{\pi}\int_{\omega_c}^\infty d\omega
\left[\cos \omega t \;  \phi_{\bf q}(0)\; \omega \; \im
F_1(-\omega^2+io,\bfq^2) \right.\\ \nonumber
&&+\sin\omega t \;  \dot \phi_{\bf q}(0)\;  \im F_1(-\omega^2+io,\bfq^2)
\\ \nonumber
&&\left.-\frac{\ddot\phi_{\bf q}(0)}{\omega}\;  \cos\omega t \; 
\im F_2(-\omega^2+io,\bfq^2)
\right]\pkt
\eea
For the consistency of the left and right hand sides and their
first derivatives w.r.t $t$ the sum rule
\be \label{sumrule1}
1=\frac{2}{\pi}\int_{\omega_c}^\infty d\omega\; 
\omega\;  \im F_1(-\omega^2+io,\bfq^2)
\ee
has to be satisfied. It follows again by considering the integral
\bea\nonumber
\frac{1}{i\pi}\oint ds \; s\;  F_1(s^2,\bfq^2)
\eea
along a closed contour around the left complex half plane.
One needs, furthermore,
\bea\nonumber
0=\frac{2}{\pi}\int_{\omega_c}^\infty d\omega\; 
 \frac{1}{\omega}\; \im F_2(-\omega^2+io,\bfq^2)
\eea
which follows from analagous considerations, using that
in this case the infinite semicircle does not contribute
as the integrand behaves as $1/s^3$ there.

We next have to consider the term proportional to
$\ddot\phi_{\bf q}(0)$ and the second time derivative of eq.(\ref{finally}).
From the renormalized equation of motion (\ref{eqm_inh_ren})
at $t=0$ we derive immediately
\be \label{sec_der}
\ddot\phi_{\bf q}(0)=-\phi_{\bf q}(0)\; \frac{\bfq^2+M^2+
\Delta\Sigmat_1(\bfq^2)}{1+\Delta\Sigmat_3(\bfq)^2}
\pkt\ee 
The second derivative of (\ref{finally}) at $t=0$ reads
\bea\nonumber
\ddot\phi_{\bf q}(0)=-
\frac{2}{\pi}\int_{\omega_c}^\infty d\omega
\left[\phi_{\bf q}(0)\; \omega^3\; \im F_1(-\omega^2+io,\bfq^2)
+\omega^2\;  \ddot\phi_{\bf q}(0)\; \im F_2(-\omega^2+io,\bfq^2)\right]\pkt
\eea
We express, on the right hand side, $\phi_{\bf q}(0)$ by
$\ddot\phi_{\bf q}(0)$, using (\ref{sec_der}). Then we can
write the sum rule as
\be\label{sumrule2}
1=\frac{2}{\pi}\int_{\omega_c}^\infty d\omega\;  \omega\; 
\im \left[\omega^2 F_1(-\omega^2+io,\bfq^2)\; \frac{1+\Delta
\Sigmat_3(\bfq^2)}{q^2+M^2+\Delta\Sigmat_1(\bfq^2)}
+F_2(-\omega^2+io,\bfq^2)\right]\pkt
\ee
The expression in the square brackets can be written explicitly as
\bea \nonumber
&&\frac{\omega^2\left(1+\Delta\Sigmat_3+\sigmat_s\right)}{
-\omega^2\left(1+\Delta\Sigmat_3+
\sigmat_s\right)+\bfq^2+M^2+\Delta\Sigmat_1}\,\frac{1+\Delta\Sigmat_1}
{\bfq^2+M^2+\Delta\Sigmat_1}\\ \nonumber
&&+\frac{\sigmat_s}
{-\omega^2\left(1+\Delta\Sigmat_3+
\sigmat_s\right)+\bfq^2+M^2+\Delta\Sigmat_1} =
\\ \nonumber
&&-\frac{1+\Delta\Sigmat_1}
{\bfq^2+M^2+\Delta\Sigmat_1}
\\ \nonumber
&&+\frac{1+\Delta\Sigmat_3+\sigmat_s}
{-\omega^2\left(1+\Delta\Sigmat_3+
\sigmat_s\right)+\bfq^2+M^2+\Delta\Sigmat_1}
\pkt\eea
Only the imaginary part if this expression occurs in the
integrand.
So the first term on the right hand side does not contribute,
and the second term is just $F_1$.  The sum rule  
(\ref{sumrule2}) reduces therefore to the first one(\ref{sumrule1}).
\end{appendix}

\begin{figure}
\centerline{ \epsfig{file=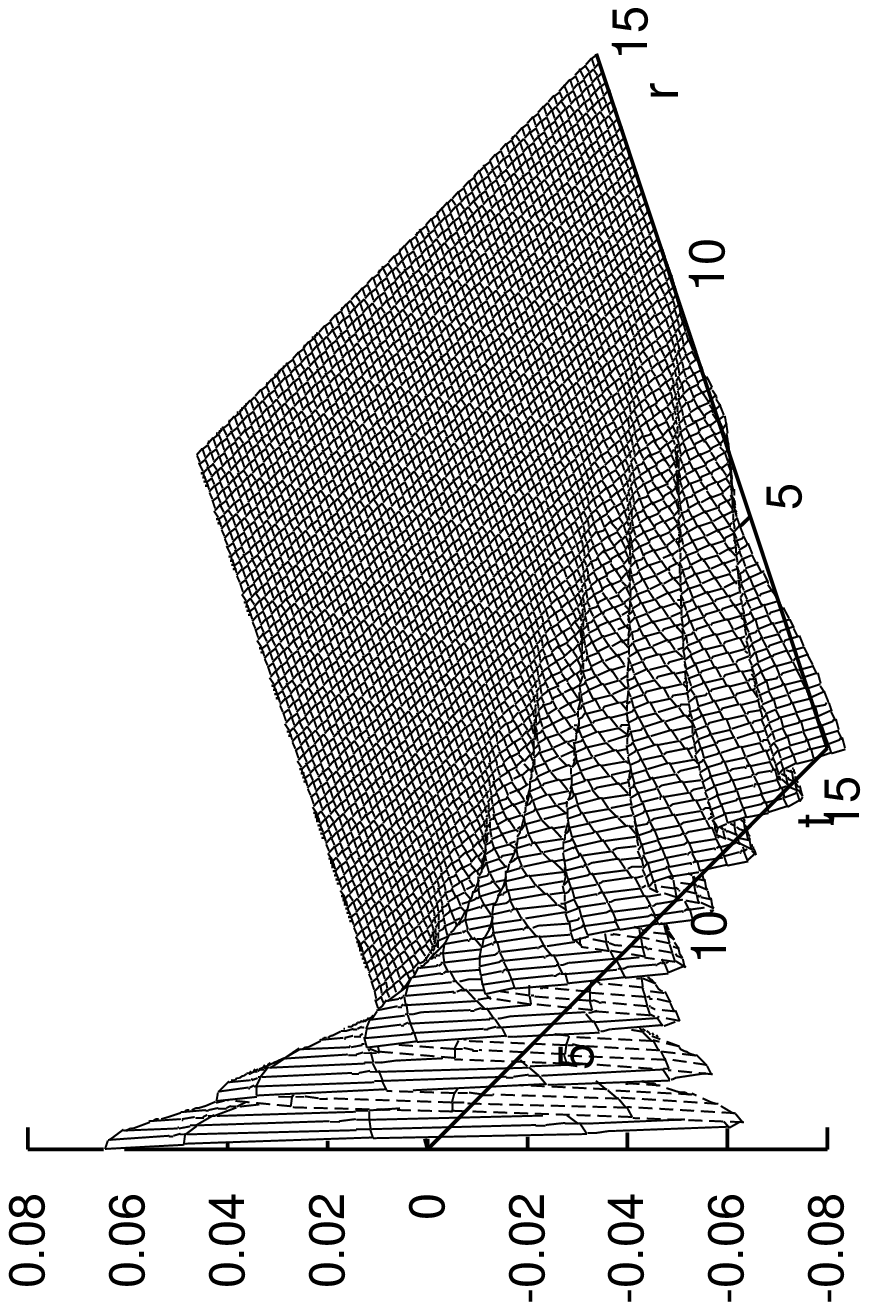,width=6in,height=6in}}
\caption{Unstable case: $M = 3~;~  m=1 ~;~g=1$ with a 
gaussian profile for the  condensate at $t=0$, given by
(\ref{gaussian}) with $R_0=1$  and normalized so that 
$\int d^3 x \phi(0,{\bf x})=1$.\label{fig1}}
\end{figure}
\begin{figure}
\centerline{ \epsfig{file=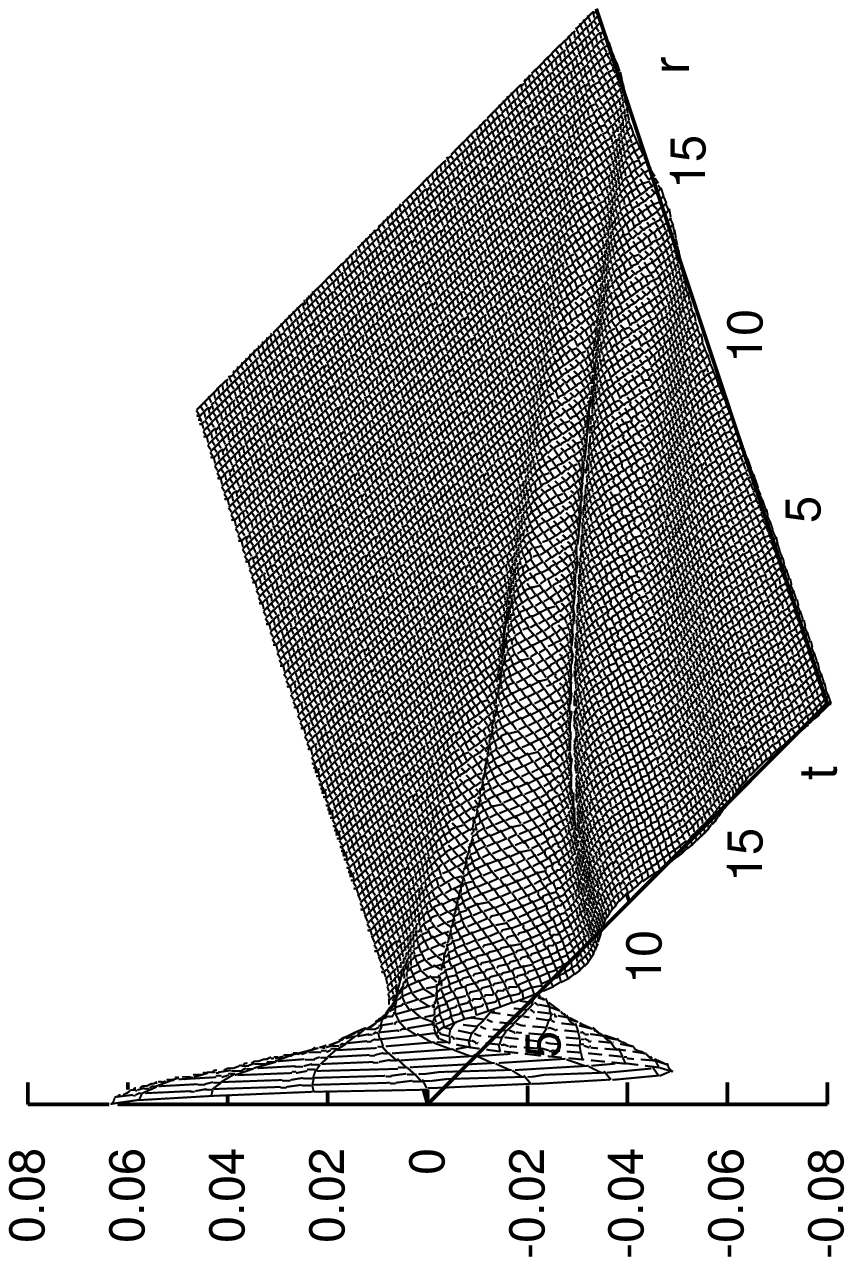,width=6in,height=6in}}
\caption{Stable case: $M = 1~;~  m=1 ~;~g=1$ with a 
gaussian profile for the  condensate at $t=0$, given by
(\ref{gaussian}) with $R_0=1$ and normalized so that 
$\int d^3 x \phi(0,{\bf x})=1$.\label{fig2}}
\end{figure}

\begin{thebibliography}{99}
\bibitem{QCD} For recent reviews on the QCD phase transitions and aspects
of relativistic heavy ion collisions see for example:  
J. W. Harris and B. Muller, Annu. Rev. Nucl. Part. Sci. {\bf 46}, 71 (1996); 
B. Muller in {\em Particle Production in Highly
Excited Matter}, 
Eds. H.H. Gutbrod and J. Rafelski, NATO ASI series B, vol. 303
(1993); 
B. Muller, {\em The Physics 
of the Quark Gluon Plasma} Lecture Notes in Physics, 
Vol. 225 (Springer-Verlag, Berlin, Heidelberg, 1985);  
K. Rajagopal in `Quark-Gluon Plasma 2', 
Ed. by R. C. Hwa, World Scientific, Singapore, 1995; 
H. Meyer-Ortmanns, Rev. of Mod. Phys. {\bf 68}, 473 (1996).
 
\bibitem{stu} E. C. G. Stueckelberg, Phys. Rev. {\bf 81}, 130 (1951).
See also, N. N. Bogoliubov and D. V. Shirkov, {\em Quantum Fields},
Benjamin, 1983.

\bibitem{largeN} 
F. Cooper, S. Habib, Y. Kluger, E. Mottola, J. P. Paz, and  
P. R. Anderson, Phys. Rev. {\bf D50}, 2848 (1994); 
F. Cooper, Y. Kluger, E. Mottola, and J.P. Paz, Phys. Rev. {\bf D51},  
2377 (1995); Y. Kluger, F. Cooper, E. Mottola, and J.P. Paz, 
Nucl. Phys. {\bf A590}, 581c (1995); M. A. Lampert, J. F. Dawson, and
F. Cooper, Phys. Rev. {\bf D54}, 2213 (1996); F. Cooper, Y. Kluger,
and E. Mottola, Phys. Rev. {\bf C54}, 3298 (1996).

\bibitem{boycosmo}
D. Boyanovsky, H. J. de Vega and R. Holman,  
{\it Nonequilibrium Dynamics of Phase Transitions: From the
Early Universe to Chiral Condensates}, Second Paris Cosmology Colloquium:
Proceedings. Eds. H. J. de Vega and N. Sanchez, World Scientific, 1995;
 D. Boyanovsky, D. Cormier, H. J. de Vega, R. Holman, S. P. Kumar,
{\it Out of Equilibrium Fields in Inflationary Dynamics: Density Fluctuations}
In the Proceedings of the 
VIth. Erice Chalonge School on Astrofundamental Physics, 
N. S\'anchez and A. Zichichi eds., Kluwer, 1998;  
 D. Boyanovsky, H. J. de Vega, R. Holman, 
{\it Erice Lectures on Inflationary Reheating}
in the Proceedings of the 
5th. Erice Chalonge School on Astrofundamental Physics, 
N. S\'anchez and A. Zichichi eds., (World Scientific, 1997), and
references therein. 
 
\bibitem{coom} F. Cooper and E. Mottola, Mod. Phys. Lett. {\bf 2}, 635
(1987);  

Phys. Rev. {\bf D36}, 3114 (1987).

\bibitem{baacke:1997}  J. Baacke, K. Heitmann, C. Patzold, 
Phys. Rev. {\bf D56} (1997) 6556.

\bibitem{ramsey}  S. A. Ramsey, B. L. Hu and  A. M. Stylianopoulos, 
Phys. Rev. {\bf D57} (1998) 6003;  S. A. Ramsey and B. L. Hu, Phys. Rev. {\bf D56} (1997) 678.

\bibitem{dcc} A. A. Anselm and M. G. Ryskin, Phys. Lett. {\bf B266}, (1991)
482;  J. - P. Blaizot and A. Krzywicki, Phys. Rev. {\bf D46}, 1992 (246);  
J. D. Bjorken, Int. J. Mod. Phys. {\bf A7}, (1992) 4189; 
J. D. Bjorken, Acta Physica Polonica {\bf B23}, (1992) 561; 
G. Amelino-Camelia, J. D. Bjorken, S. E. Larsson, 
Phys. Rev. {\bf D56} (1997) 6942;  
J. D. Bjorken, Acta Phys.Polon. {\bf B28} (1997) 2773; 
A. Anselm, Phys. Lett. {\bf B217}, 169 (1989). 
 K. Rajagopal and F. Wilczek, Nucl. Phys. {\bf B399}, 395 (1993);
K. Rajagopal and F. Wilczek, Nucl. Phys. {\bf B404}, 577 (1993). 
 S. Gavin, A. Gocksch and R. D. Pisarski, 
Phys. Rev. Lett, {\bf 72}, 2143 (1994); 
S. Gavin and B. Muller, Phys. Lett. {\bf B329}, 486 (1994);
 Z. Huang and X.-N. Wang, Phys. Rev. {\bf D49}, 4335 (1994); 
Z. Huang, M. Suzuki and X-N. Wang, Phys. Rev. {\bf D50}, 2277 (1994); 
 Z. Huang and M. Suzuki, Phys. Rev. {\bf D53}, 891 (1996); 
M. Asakawa, Z. Huang and X. N. Wang, Phys. Rev. Lett. {\bf 74}, 3126 (1995); 
D. Boyanovsky, H. J. de Vega and R. Holman, Phys. Rev. {\bf D51}, (1995) 734; 
F. Cooper,  Y. Kluger, E. Mottola and J. P.
Paz. Phys. Rev. {\bf D51}, (1995) 2377; 
Y. Kluger, F. Cooper, E. Mottola, J. P. Paz and A.
Kovner, Nucl. Phys. {\bf A590},(1995) 581; 
J. Randrup, Nucl.Phys. {\bf A616} (1997) 531; 
J. Randrup, Phys. Rev. Lett. {\bf 77} (1996) 1226; 
H. Minakata and B. Muller, Phys. Lett. {\bf B377}, 135 (1996); 
M. Asakawa, H. Minakata and B. Muller, Phys. Rev. {\bf D58}, 094011 (1998);
 D. Boyanovsky, H. J. de Vega, R. Holman and S. Prem Kumar,
Phys. Rev.  {\bf D56}, 3929 (1997); {\it ibid} 5233 (1997).

\bibitem{boydcc} 
D. Boyanovsky, H. J. de Vega, and R. Holman, Phys. Rev. 
{\bf D51}, 734 (1995); 
D. Boyanovsky, D.-S. Lee and A. Singh, Phys. Rev. {\bf D48}, 800 (1993);  
D. Boyanovsky, H. J. de Vega, R. Holman, S. Prem Kumar, R. D. Pisarski, 
Phys. Rev. {\bf D57} (1998) 3653.

\bibitem{vautherin} Y. Tsue, D. Vautherin, T. Matsui, hep-ph/9812254; 
D. Vautherin and T. Matsui, Phys.Lett. {\bf B437}, 137 (1998).

\bibitem{boycoopveg} D. Boyanovsky, F. Cooper, H. J. de Vega and  P. Sodano,   
Phys. Rev. {\bf D58},  (1998) 025007.

 \bibitem{kluger}  S. Habib, Y. Kluger, and E. Mottola, Phys. Rev. {\bf D55},
6471 (1997); Y. Kluger, J. M. Eisenberg, B. Svetitsky, F. Cooper
and E. Mottola,
Phys. Rev. {\bf D45}, 4659 (1992); Phys. Rev. Lett. {\bf 67}, 2427 (1991).

\bibitem{baackegauge} J. Baacke, K. Heitmann and C. Paetzold, 
Phys. Rev. {\bf D55} (1997) 7815.

\bibitem{aarts} G. Aarts and J. Smit,  hep-ph/9902231, hep-ph/9812413,
hep-ph/9809340 and hep-ph/9906538.  

\bibitem{boyinho}  D. Boyanovsky, M. D'Attanasio, H.J. de Vega and
R. Holman, Phys. Rev. {\bf D54}, 1748 (1996). 

\bibitem{boyareheat} D. Boyanovsky, M. D'Attanasio, H. J. de Vega, R. Holman,
and D.-S. Lee, Phys. Rev. {\bf D52}, 6805 (1995); {\it New Aspects of
Reheating}, String Gravity and Physics at the Planck Energy Scale:
Proceedings of the NATO ASI at Erice, Italy, p.451, 
Eds. N. Sanchez  and A. Zichichi, Kluwer, 1996.

\bibitem{Baacke:1998a} J. Baacke, K. Heitmann, and
C. P\"atzold, Phys. Rev.  {\bf D57}, 6398 (1998); {\em ibid} 6406.

\bibitem{Baacke:1998c} J. Baacke, K. Heitmann, and
C. P\"atzold, Phys. Rev. {\bf D58} (1998) 125013.

\bibitem{sym} K. Symanzik, Nucl. Phys. {\bf B190}, 1 (1981).

\bibitem{ruso} V. P. Maslov and O. Yu. Shvedov,
Theor. Math. Phys. {\bf 114}, 184 (1998).

\bibitem{ctp}J. Schwinger, J. Math. Phys. {\bf 2}, 407 (1961); 
K. T. Mahanthappa, Phys. Rev.  {\bf 126}, 329 (1962); 
P. M. Bakshi and K. T. Mahanthappa, J. Math. Phys.  {\bf 41}, 12 (1963)
; A. Niemi and G. Semenoff, Ann. of Phys. (NY) {\bf 152}, 105 (1984); 
N. P. Landsmann and C. G.  van Weert, Phys. Rep. {\bf 145}, 141 (1987); 
E. Calzetta and B. L. Hu, Phys. Rev.  {\bf D41}, 495 (1990); {\em ibid}  
{\bf D37}, 2838  (1990); J. P. Paz, Phys. Rev. {\bf D41}, 1054 (1990);
{\em ibid} {\bf D42}, 529(1990). 

\bibitem{keldysh} L. V. Keldysh, JETP  {\bf 20}, 1018 (1965); 
K. Chou, Z. Su, B. Hao And L. Yu, Phys. Rep.  {\bf 118}, 1 (1985). 

\bibitem{boyRG} D. Boyanovsky, H. J. de Vega, R. Holman, M. Simionato, 
hep-ph/9809346, to appear in Phys. Rev. {\bf D}. 

\bibitem{shang} S.-Y. Wang, D. Boyanovsky, H. J. de Vega, 
D.-S. Lee and Y. J. Ng, hep-ph/9902218; D. Boyanovsky,
H. J. de Vega, D.-S. Lee, Y. J. Ng and S.-Y. Wang, Phys. Rev. {\bf D59}, 105001 (1999). 
\end{thebibliography}
\end{document}